\documentclass[11pt]{article}

\usepackage{fancyhdr}

\usepackage{geometry}

\usepackage{bm}
\usepackage{graphicx}
\usepackage{amsmath}
\usepackage{natbib}
\usepackage[english]{babel}
\usepackage{float}
\usepackage{etoolbox}
\usepackage{multirow}
\usepackage{url}

\usepackage{lscape}
\usepackage[english]{babel}

\usepackage{bm}
\usepackage{bbm}
\usepackage{tikz}
\usepackage{relsize}

\usepackage{amssymb}
\usepackage{amsthm}
\usepackage{mathtools}
\usepackage{enumerate}
\usepackage{xcolor}
\usepackage{longtable}
\usepackage[ruled,vlined,linesnumbered,lined,boxed,commentsnumbered]{algorithm2e}
\usetikzlibrary{positioning}
\usetikzlibrary{graphs} % LATEX and plain TEX
\usetikzlibrary[graphs]

\usepackage[normalem]{ulem}

\definecolor{gray1}{rgb}{0.90, 0.90, 0.90}
\definecolor{gray2}{rgb}{0.70, 0.70, 0.70}
\definecolor{gray3}{rgb}{0.45, 0.45, 0.45}
\definecolor{gray4}{rgb}{0.25, 0.25, 0.25}

\newcommand{\bc}{\begin{center}}
\newcommand{\ec}{\end{center}}

\newcommand{\bit}{\begin{itemize}}
\newcommand{\eit}{\end{itemize}}

\newcommand{\be}{\begin{eqnarray*}}
\newcommand{\ee}{\end{eqnarray*}}
\newcommand{\ben}{\begin{eqnarray}}
\newcommand{\een}{\end{eqnarray}}
\newcommand{\g}{\,\vert\,}

\newcommand{\G}{\mathcal{G}}

\newcommand{\X}{\mathcal{X}}

\newcommand{\bA}{\bm{A}}

\newcommand{\bX}{\bm{X}}

\newcommand{\ba}{\bm{a}}

\newcommand{\bx}{\bm{x}}

\newcommand{\btheta}{\bm{\theta}}

\newcommand{\black}{\color{black}}

\usepackage{tikz}
\usetikzlibrary{positioning}

\newtheorem{prop}{Proposition}[section]

\usepackage{setspace}

\SetKwInput{KwInput}{Input}
\SetKwInput{KwOutput}{Output}

\usepackage{authblk}

\begin{document}

%\title{Clustering multivariate categorical data: \\a graphical model-based approach}
\title{Graphical model-based clustering of categorical data}
\author[1]{Laura Ferrini \thanks{laura.ferrini@unicatt.it}}
\author[2]{Federico Castelletti \thanks{federico.castelletti@unicatt.it}}
\affil[1,2]{Department of Statistical Sciences, Universit\`{a} Cattolica del Sacro Cuore, Milan}
\affil[1]{Department of Economics, Management and Statistics, Universit\`{a} degli Studi di Milano-Bicocca, Milan}

\date{}

\maketitle

%\maketitle
%\noindent

\begin{abstract} % max 200 words

Clustering multivariate data is a pervasive task in many applied problems, particularly in social studies and life science.
Model-based approaches to clustering rely on mixture models, where each mixture component corresponds to the kernel of a distribution characterizing a latent sub-group. Current methods developed within this framework employ multivariate distributions built under the assumption of independence among variables given the cluster allocation. Accordingly, possible dependence structures characterizing differences across groups are not directly accounted for during the clustering process.
In this paper we consider multivariate categorical data %multivariate data consisting of $q$ categorical variables collected over $n$ statistical units,
and introduce a model-based clustering method which employs graphical models as a tool to encode dependencies between variables. Specifically, we consider a Dirichlet Process mixture of categorical graphical models, which clusters individuals into groups that are homogeneous in terms of dependence (graphical) structure and allied parameters. We provide full Bayesian inference for the model and develop a Markov chain Monte Carlo scheme for posterior analysis. Our method is evaluated through simulations and applied to real case studies, including the analysis of genomic data %of patients affected by cystic fibrosis,
and voting records. Results reveal the merits of a graphical model-based clustering, in comparison with approaches that do not explicitly account for dependencies in the multivariate distribution of variables.

\vspace{0.7cm}
\noindent
Keywords: Bayesian inference; Dirichlet Process prior; heterogeneous data; Markov chain Monte Carlo.

\end{abstract}

\section{Introduction}
\label{sec:Introduction}

%Letteratura su clustering per dati categorici

Clustering represents a challenging problem in many applied contexts including social sciences, biology, and medicine. In this paper we consider multivariate data consisting of $q$ categorical variables collected over $n$ statistical units that we aim to cluster into homogeneous groups.
To address this task, different methodologies have been proposed in the literature, including both distance- and model-based approaches. Among the first, \textit{k-modes} \citep{kmodes} implements a modified version of the popular \textit{k-means} algorithm \citep{MacQueen:1967}
and adopts a dissimilarity measure between modes of categorical variables to cluster observations into homogeneous groups. 
Model-based clustering instead relies on mixture models, where each observation is assumed to arise from one of (potentially infinitely many) latent components. In this framework, a typical assumption is that variables are (conditionally) independent given cluster memberships. As an example, the poLCA (polytomous variable Latent Class Analysis) method \citep{linzer} employs a mixture model where each mixture component is a joint distribution factorizing into the product of univariate categorical kernels. Each component is therefore parameterized by its own set of marginal probabilities corresponding to the levels of a categorical variable. Inference on such parameters is carried out \textit{via} an EM algorithm, and the number of components is chosen using model selection criteria. A Bayesian formulation of the model requires the specification of prior distributions for each mixture-component parameter and possibly the number of mixture components itself; see in particular \citet{White:Murphy:BayesLCA}.

%%%%%%%%%%%%%%%
%Alternatively, the latter can be inferred \emph{a posteriori} by comparing models with different number of components through information criteria.
%%%%%%%%%%%%%%%
%This basic mixture model is also extended to latent class regression; the latter allows for the inclusion of covariates that can help in the prediction of individual class memberships.
%Importantly however, none of these methods accounts for possible dependence relationships between variables in the underlying multivariate statistical model, which instead represents a peculiar feature of our methodology.
%%%%%%%%%%%%%%%
%and it is also crucial for causal inference purposes.
%%%%%%%%%%%%%%%
%It is proposed that a latent variable, following a mixture of Gaussian distributions, generates the observed data of mixed type. The observed data may be any combination of continuous, binary, ordinal or nominal variables. clustMD employs a parsimonious covariance structure for the latent variables, leading to a suite of six clustering models that vary in complexity and provide an elegant and unified approach to clustering mixed data. An expectation maximisation (EM) algorithm is used to estimate clustMD
%%%%%%%%%%%%%%%
%More recently, \citet{McParland:Gormley:2015} propose a method for clustering mixed (i.e.~continuous, binary, ordinal or nominal) data which is instead based on a latent multivariate Gaussian distribution and therefore accounts for a covariance structure between variables at the latent level.
%%%%%%%%%%%%%%%

The Bayesian nonparametric framework offers a rich variety of parameter prior distributions whose discreteness induces a partition of the observations and thus allows for clustering too. Among these, Dirichlet Process mixtures \citep{Antoniak:1974}
represent a well-established class of models with infinite number of components; inference can be carried out by efficient algorithms based on  Markov chain Monte Carlo methods; see in particular \citet{Neal:2000}. \black
Finite mixtures have also attracted considerable interest in the last years. In this regard, \citet{Argiento:De:Iorio:2022} presented a broad class of mixing measures obtained by normalization of finite point processes, which includes the finite Dirichlet Process as a special case. Particularly, this framework was employed by \citet{argientopaci} for clustering multivariate categorical data. The Authors considered, as a probability mass function for a categorical variable, the so-called Hamming distribution, which is parameterized by a location and a dispersion parameter.
The joint distribution of $q$ categorical variables is defined under the assumption of conditional independence and thus factorizes as the product of univariate Hamming distributions. It is then employed as the kernel of a finite mixture model, for which the Authors provide full Bayesian inference and a related Markov chain Monte Carlo algorithm for posterior analysis.
More recently, \citet{Malsiner:et:al:2024} relaxed the conditional independence assumption between variables by proposing a mixture of finite mixtures model with a two-layer structure. In the hierarchical formulation of the model, an upper layer governs the cluster allocation, whereas a lower layer offers a highly flexible representation of the cluster-specific distributions, thereby accommodating possible dependencies among variables within clusters.

In this work, we pursue a different route to relax conditional independence, motivated by settings where an explicit dependence structure among variables can characterize data heterogeneity. We base our methodology on graphical models which offer a powerful tool for learning dependence structures between variables from the data. Specifically, we consider an infinite mixture of categorical graphical models based on a Dirichlet Process prior having as baseline measure a joint distribution over graphs and related parameters. As such, our model clusters individuals into groups that are homogeneous in terms of graphical (dependence) structure and model parameters, which therefore both inform the clustering process.

%We employ as kernel distributions within a mixture model, the multivariate distribution of a categorical graphical model. This choice introduces an additional, interpretable parameter --namely, the graph itself-- that captures within-cluster dependencies and simultaneously contributes to the clustering mechanism. Concerning the mixing measure, we consider a Dirichlet Process prior that allows for ties across model parameters and graphs. 

%%%% SE SI VUOLE AGGIUNGERE LETTERATURA SI PUO GUARDARE UN PO COSA DICONO WALLI ETC ETC  %%%%%%%%

%Importantly however, none of the methods discussed above accounts for possible dependence relationships between variables in the underlying multivariate statistical model, which represents a peculiar feature of our methodology for clustering, and it is also crucial for causal inference purposes.
%and structure learning
%and in turn heterogeneous causal inference.

% va asciugata la parte dopo; troppi DAG
Structure learning of categorical graphical models has a long tradition. The most popular frequentist method is the PC algorithm of \citet{Spir:Glym:Sche:2000}, a constraint-based approach which applies sequences of conditional independence tests to recover the structure of a Completed Partially Directed Acyclic Graph (CPDAG); see also \citet{Kalisch:Buhlmann:2007}. A score-based methodology is instead at the basis of the Hill Climbing (HC) algorithm \citep{Russell:2016} which for the case of categorical data adopts the Bayesian Dirichlet equivalent uniform (BDeu) score of \citet{Heckerman:et:al:1995}
to traverse through optimization methods the space of Directed Acyclic Graphs (DAGs); see also \citet{Scutari:Denis:R:2021} for computational details.
Moving to the Bayesian framework, an early work for structure learning of categorical DAGs is \cite{Madigan:York:1995}, who also propose a Markov Chain Monte Carlo (MCMC) strategy to approximate a posterior distribution over the space of DAGs.
More recently, \cite{Castelletti:Peluso:2021:CSDA} instead proposed a method for structure learning of CPDAGs, based on an MCMC scheme designed over the CPDAG space.
The main challenge in Bayesian graphical modelling is represented by prior elicitation for model parameters that must reflect the conditional independencies encoded by the graph. With regard to DAGs, \citet{Geiger:Heckerman:1997} developed a constructive elicitation procedure which applies to categorical data distributions with Dirichlet priors. For undirected (decomposable) graphs instead, \citet{Dawid:Laur:1993} introduced a class of Hyper Markov laws, that are specialized to the categorical setting through the so-called Hyper-Dirichlet priors, and are also implemented in this work.

The rest of the paper is organized as follows. In Section \ref{sec:background} we provide some relevant background material on  Bayesian inference for categorical data and decomposable graphical models. In Section \ref{sec:mixtures} we introduce our Dirchlet Process mixture model, for which we detail the construction of the baseline over graphs and parameters. Section \ref{sec:Computational:details} introduces our MCMC scheme for posterior inference. This is implemented on simulated data in Section \ref{sec:simulation:study}, which includes comparisons of our methodology with alternative approaches for clustering categorical data. Section \ref{sec:Application} is instead devoted to real data analyses, including voting records and genomic data of patients affected by cystic fibrosis. Finally, Section \ref{sec:Discussion} offers a brief discussion together with possible extensions of our method. A few technical results are included in the Supplementary Material.

% aggiungere che metodi che non considerano la dipendenza tenderanno a sovrastimare il numero di cluster  + si puo fare vedere??? si ma devo modificare le simulazioni sui metodi di confronto e FRuwischnatter

\section{Background}
\label{sec:background}

In this section we provide some supporting background material on categorical data analysis and graphical models.

\subsection{Categorical data distributions}
\label{subsec:Categorical:data}

Let $X=(X_1,\dots,X_q)^\top$ be a vector of categorical variables, with each $X_j \in \X_j$, the set of its levels and $l_j=|\X_j|$.
Let also $\X=\times_{j=1}^{q}\X_j$ be the product space generated by the $q$ variables, $x=(x_1,\dots,x_q)$ a generic element of $\X$.
A dataset consisting of $n$ multivariate observations from $X_1,\dots, X_q$ can be represented as an $(n,q)$ matrix $\bX$ with rows $\bx^{(1)},\dots,\bx^{(n)}$, where $\bx^{(i)}=\big(x_{1}^{(i)}, \dots, x_{q}^{(i)}\big)$ and each $\bx^{(i)}\in\X$, $i=1,\dots,n$.
%An instance of list for $n=10$ individuals and $q=3$ variables, $X_1$ (Country), $X_2$ (Gender), $X_3$ (Employment), where $\X_1=\{\textnormal{IT},\textnormal{ES}\}$, $\X_2=\{\textnormal{M},\textnormal{F}\}$, $\X_3=\{\textnormal{yes},\textnormal{no}\}$ is reported in Table \ref{tab:list}. In this case we have
%\be
%\X &=& \textnormal{\{(IT, M, yes), (IT, M, no), (IT, F, yes), (IT, F, no),} \nonumber \\
%&& \, \, \, \textnormal{(ES, M, yes), (ES, M, no), (ES, F, yes), (ES, F, no)\}}.
%\ee
%
%\begin{figure}[H]
%	\begin{center}
%		\begin{tabular}{cccc}
%			\hline
%			$i$ & \textnormal{Country} & \textnormal{Gender} & \textnormal{Employment} \\
%			\hline
%			%\multirow{5}{*}{$...$}
%			$1$ & \textnormal{IT} & \textnormal{M} & \textnormal{yes} \\
%			$2$ & \textnormal{ES} & \textnormal{F} & \textnormal{no} \\
%			\vdots & \vdots & \vdots & \vdots \\
%			$10$ & \textnormal{ES} & \textnormal{M} & \textnormal{yes} \\
%			\hline
%		\end{tabular}
%	\end{center}
%	\caption{A data set with $n=10$ observations and $q=3$ variables represented as a $(10,3)$ list.}
%	\label{tab:list}
%\end{figure}
%
The set $\X$ can be arranged as a $q$-way contingency table, whose cells correspond to the elements $x\in\X$.
Consider now, for each $x\in\X$, the joint probability $\theta_x=\Pr \left\{X=x\right\}=\Pr \left\{X_1=x_1, \dots, X_q=x_q\right\}$, such that $0 \le \theta_x \le 1$, and $\sum_{x\in\X}\theta_x = 1$. The ensuing collection $\{\theta_x, x\in\X\}$ can be arranged in a $q$-way table of probabilities which we denote by $\btheta$.
%
%For instance, given $S=\{1,2\}$ in the previous example we have
%\be
%\X_S &=& \textnormal{\{(IT, M), (IT, F), (ES, M), (ES, F)\}}
%\ee
%and the marginal table represented in Figure \ref{tab:marg:table}.
%The allied marginal probabilities $\pi(x_S)$ can be obtained from the elements $\pi(x)$ in $\bPi$ as
%\ben
%\pi(x_S) &= &\sum_{x \in \X}\pi(x)\mathbbm{1}\{x(S)=x_S\},
%\een
%where $x(S)$ denotes the sub-vector of $x$ with elements in the set $S$ and $\mathbbm{1}(\cdot)$ is the indicator function.
%The marginal probabilities $\pi(x_s), x_s \in \X_S$ can be then collected in the $|S|$-dimensional table $\bPi_S$.
%
%\begin{figure}[H]
%	\begin{center}
%		\begin{tabular}{cccc}
%			\hline
%			&& \multicolumn{2}{c}{Country} \\
%			& & IT & ES \\
%			\hline
%			\multirow{2}{*}{Gender}
%			& M & (M, IT) & (M, ES) \\ 
%			& F & (F, IT) & (F, ES) \\ 
%			\hline
%		\end{tabular}
%	\end{center}
%	\caption{A marginal table with two variables.}
%	\label{tab:marg:table}
%\end{figure}
%
Moreover, for each $x \in \X$ we can define the count
\be
n_x \,=\, \sum_{i=1}^{n}\mathbbm{1}\big\{\bx^{(i)}=x\big\},
\ee
corresponding to the number of observations (rows) in $\bX$ that are equal to $x\in\X$. 
For any subset $S\subseteq V = \{1,\dots,q\}$, let now $X_S$ be the sub-vector of $X$ with variables in $S$ and $\X_S=\times_{j\in S}\X_j$ the corresponding marginal table, with elements $x_S\in\X_S$. Let also $\bx^{(i)}_S$ be the sub-vector of $\bx^{(i)}$ with components indexed by $S$.
The marginal count
\be
n^S_{x_S} \,=\, \sum_{i=1}^{n}\mathbbm{1}\left\{\bx^{(i)}_S=x_S\right\}
%&=& \sum_{x \in \X} n(x)\mathbbm{1}\big\{x(S)=x_S\big\}.
\ee
corresponds to the number of observations presenting configuration $x_S \in \X_S$ of variables belonging to set $S$.
Finally, for each $x_S \in \X_S$, we let $\theta_{x_S}^S=\Pr\{X_S = x_s\}$ be the corresponding marginal probability for variables in $S$. As before, the collection of probabilities $\{\theta^S_{x_S},x_S\in\X_S\}$ can be arranged in an $|S|$-way marginal table $\btheta_S$.
%magari definire direttamente il vettore dei thetaS
%
Following the notation above, the sampling distribution of $\bx^{(i)}$ can be written as
\be
p\big(\bx^{(i)}\g\btheta\big)
%&=& \Pr\left\{X_1 = x_1^{(i)}, \dots, X_q = x_q^{(i)}\right\} \nonumber \\
\,=\,
\prod_{x \in \X}\left\{\theta_x\right\}^{\mathbbm{1}\left\{\bx^{(i)}= x\right\}}
\ee
and the likelihood function for $n$ i.i.d. observations is given by
\ben
\label{eq:like}
p\big(\bX\g\btheta\big) %&=& \prod_{i=1}^{n}p\big(\bx^{(i)}\g\btheta\big) \nonumber \\
%\,=\, \prod_{i=1}^{n} \left\{ \prod_{x \in \X}\left\{\theta_x\right\}^{\mathbbm{1}\left\{\bx^{(i)}=x\right\}} \right\}
\,=\, \prod_{x \in \X}\left\{\theta_x\right\}^{n_x},
\een
which depends on the raw data $\bX$ through the counts $n_x$.
%which thus represent the sufficient statistic.
%
Moreover, for any $S\subseteq V$, the marginal sampling distribution restricted to variables in $S$ is
\be
p\big(\bx^{(i)}_S\g\btheta_S\big) \,=\, \prod_{x_S \in \X_S}\left\{\theta^S_{x_S}\right\}^{\mathbbm{1}\left\{\bx^{(i)}_S= x_S\right\}}.
\ee
Finally, if we let $\bX_S$ be the sub-matrix of $\bX$ with columns indexed by $S$, the likelihood is
\ben
\label{eq:likelihood:subset}
p\big(\bX_S\g\btheta_S\big) %&=& \prod_{i=1}^{n}p\big(\bx^{(i)}_S\g\btheta_S\big) \nonumber \\
%\,=\, \prod_{i=1}^{n} \left\{\prod_{x_S \in \X_S}\left\{\theta^S_{x_S}\right\}^{\mathbbm{1}\left\{\bx^{(i)}_S=x_S\right\}} \right\}
\,=\, \prod_{x_S \in \X_S}\left\{\theta^S_{x_S}\right\}^{n^S_{x_S}}.
\een

%%%%%%%%%%%%%%%%%%%%%%%%%%%%%

\subsection{Graphical models}
\label{subsec:Graphical:models}

\begin{figure}[H]
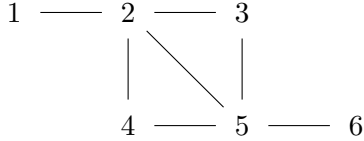

	\begin{center}
		\begin{tabular}{c}
			{\normalsize
				\tikz \graph [no placement, math nodes, nodes={circle}]
				{
					1[x=0,y=0] -- 2[x=1.5,y=-0],
					2[x=1.5,y=-0] -- 3[x=3,y=-0],
					2[x=1.5,y=-0] -- 4[x=1.5,y=-1.5],
					4[x=1.5,y=-1.5] -- 5[x=3,y=-1.5],
					3[x=3,y=-0] -- 5[x=3,y=-1.5],
					2[x=1.5,y=-0] -- 5[x=3,y=-1.5],
					5[x=3,y=-1.5] -- 6[x=4.5,y=-1.5]};
			}
		\end{tabular}
	\end{center}
	\caption{\small A decomposable graph $\G$ on the set of vertices $V=\{1,\dots,6\}$. $\G$ has the set of cliques $\mathcal{C}=\{C_1,C_2,C_3,C_4\}$ with $C_1=\{1,2\},C_2=\{2,3,5\},C_3=\{2,4,5\},C_4=\{5,6\}$, and set of separators $\mathcal{S}=\{S_2,S_3,S_4\}$, with $S_2=\{2\},S_3=\{2,5\},S_3=\{5\}$.}
	\label{fig:decomposable}
\end{figure}

Let $\G=(V,E)$ be an Undirected Graph (UG) consisting of a set of vertices (nodes) $V=\{1,\dots,q\}$ and a set of edges $E \subseteq V \times V$, such that if $(u,v) \in E$ then also $(v,u)\in E$. Graph $\G$ can be represented through a $(q,q)$ adjacency matrix $\bA^\G$, with $(u,v)$-element $\bA^\G_{u,v}$, and such that $\bA^\G_{u,v}=\bA^\G_{v,u}=1$ if and only if $\G$ contains the undirected edge $u - v$, namely $(u,v),(v,u)\in E$.
For any subset $S\subseteq V$ we let $\G_{S}=(S,E_{S})$, with $E_{S}=\{(u,v) \in E \g u,v \in S\}$, be the sub-graph of $\G$ induced by $S$.
In an UG $\G$, two vertices $u,v$ are adjacent if they are connected by an edge. A (sub)graph is complete if its vertices are all adjacent.
%In addition, we call $u$ a \textit{neighbor} of $v$ if $u - v$ is in $\G$ and denote the neighbor set of $v$ as $\neigh_{\G}(v)$; the common neighbor set of $u$ and $v$ is then $\neigh_{\G}(u,v)=\neigh_{\G}(u) \cap \neigh_{\G}(v)$.
%We say instead that $u$ is a \textit{parent} of $v$ if $u\rightarrow v$ is in $\G$; conversely, we say that $v$ is a son of $u$. The set of all parents of $u$ in $\G$ is then denoted by $\pa_{\G}(u)$.
Moreover, a sequence of nodes $(v_0,v_1,\dots,v_k)$ such that $v_0=v_k$ and $v_{j-1} - v_j$ for all $j=1,\dots,k$ is called a \textit{cycle}.
%A cycle is directed (undirected) if it contains only directed (undirected) edges; conversely we call it a partially-directed cycle. A graph with only directed edges is called a Directed Acyclic Graph (DAG) if it does not contain cycles.
%
An UG is \emph{decomposable} if every cycle of length $l\ge 4$ has a \textit{chord}, that is two nonconsecutive vertices that are adjacent. For a decomposable graph $\G$, a complete subset that is maximal with respect to inclusion is called a \textit{clique}. Let $\mathcal{C} = \{C_1,\dots, C_K\}$ be the set of cliques of $\G$. Let also $H_k = C_1 \cup \dots \cup C_k$, for $k=2,\dots,K$ and $H_1=\O$. Then, the set of \emph{separators} $\mathcal{S} = \{S_2,\dots,S_K\}$ has elements $S_k = C_k \cap H_{k-1}$; see also \citep[p.18]{Laur:1996} and Figure \ref{fig:decomposable} for an example.
%It can be shown \citep[p.18]{Laur:1996} that each decomposable graph can be uniquely represented by its set of cliques and separators.
%Most importantly, for each decomposable graph one can obtain a \textit{perfect numbering} of its vertices \citep{Laur:1996} and then a \textit{perfect directed version} $\G^{<}$ by directing its edges from lower to higher numbered vertices
%; see also Figure \ref{fig:dec:perfect}.
%
\vspace{0.4cm}
\noindent
Under a decomposable UG $\G$, the likelihood of a categorical model whose $q$ variables are associated with the nodes of $\G$ factorizes as
\ben
\label{eq:likelihood:decomposable}
p\big(\bX\g \btheta, \G\big) = \frac{\prod_{C \in \mathcal{C}} p\big(\bX_C\g\btheta_C\big)}
{\prod_{S \in \mathcal{S}} p\big(\bX_S\g\btheta_S\big)},
\een
\black
with each term as in \eqref{eq:likelihood:subset}, and where we emphasize the dependence on graph $\G$ in the left-hand side of the equation, besides the graph-dependent parameter $\btheta$. We refer the reader to \citet{Laur:1996} for further notation on categorical decomposable graphical models.

\section{Mixtures of categorical graphical models}
\label{sec:mixtures}

In this section we introduce our mixture model for the analysis of multivariate categorical data. This is based on a Dirichlet Process (DP) mixture of categorical UG models which can be written according to the following hierarchical structure:
\begin{eqnarray} \label{dpmixture}
	\boldsymbol{x}^{(i)} \g \boldsymbol{\theta}_i , \mathcal{G}_i  &\sim& p(\boldsymbol{x}^{(i)} \g \boldsymbol{\theta}_i, \mathcal{G}_i) \nonumber \\
	(\boldsymbol{\theta}_i, \mathcal{G}_i) \g H  &\sim& H \\
	H &\sim& DP(M_0, \alpha).\nonumber 
\end{eqnarray}

In particular, $DP(M_0, \alpha)$ denotes a Dirichlet Process prior with concentration parameter $\alpha > 0$ and baseline measure $M_0$ \citep{Ferguson:1973}. The baseline $M_0$ represents the prior distribution over the space of graph-parameter pairs $(\boldsymbol{\theta}, \mathcal{G})$, while $\alpha$ controls the concentration of the random measure $H$ around $M_0$. As an interpretation, larger values of $\alpha$ lead to higher dispersion and thus a higher expected number of distinct clusters in the data; smaller values instead encourage stronger sharing of atoms and hence a smaller number of clusters.

A key theoretical property of the DP is that the random probability measure $H$ is almost surely discrete, i.e.
$
H = \sum_{l=1}^{\infty} w_l \, \delta_{(\boldsymbol{\theta}_l, \mathcal{G}_l)},
$
where $\{w_l\}_{l\ge 1}$ are non-negative random weights summing to one, $\delta_{(\boldsymbol{\theta}_l, \mathcal{G}_l)}$ denotes a point mass located at $(\boldsymbol{\theta}_l, \mathcal{G}_l)$, and the atoms $\{(\boldsymbol{\theta}_l, \mathcal{G}_l)\}_{l\ge 1}$ are independent draws from the baseline measure $M_0$. This discreteness implies that, with positive probability, multiple observations will share exactly the same values of $(\boldsymbol{\theta}_i, \mathcal{G}_i)$.
As a consequence, model \eqref{dpmixture} induces a random partition of the observations $\boldsymbol{x}^{(1)}, \dots, \boldsymbol{x}^{(n)}$ into clusters. In our context, observations assigned to the same cluster share the same atom $(\boldsymbol{\theta}_l, \mathcal{G}_l)$ and are therefore characterized by a common graphical (dependence) structure between variables and related graph-parameter.
%\gray In other words, within each cluster the conditional independence relationships among variables are encoded by the same graph $\mathcal{G}_l$, while $\boldsymbol{\theta}_l$ governs the distribution compatible with that graph. \black
The resulting model can thus be viewed as a nonparametric mixture of categorical graphical models, where both the clustering partition and the graphical structure characterizing each cluster are learned from the data in a fully Bayesian fashion. %Moreover, the expected number of clusters is controlled by $\alpha$: each observation $\boldsymbol{x}^{(i)}$ is associated with its own $(\boldsymbol{\theta}_i, \D_i)$-parameter as $\alpha \rightarrow \infty$; on the contrary, if $\alpha \rightarrow 0$, then all observations are assigned to the same cluster, leading to a standard categorical DAG model \citep{castelletti2023joint}; see also \cite{muller2013dirichlet} for related properties of the DP prior.

Let now $K \leq n$ be the number of unique values among $\{(\boldsymbol{\theta}_i, \mathcal{G}_i)\}_{i=1}^n$, and $\{\xi_i\}_{i=1}^n$ a sequence of cluster indicator variables such that $\xi_i \in \{1, \dots, K\}$ and $(\boldsymbol{\theta}_i, \mathcal{G}_i) = (\boldsymbol{\theta}_{\xi_i}, \mathcal{G}_{\xi_i})$.
Conditionally on $\{\xi_i\}_{i=1}^n$, observations are i.i.d.~within each cluster, so that the likelihood can be written as
\begin{equation}\label{eq:lik_clustering}
	\begin{aligned}
		p\left(\boldsymbol{X} \g \{\xi_i\}_{i=1}^n, \{\btheta_i\}_{i=1}^n, \{\mathcal{G}_i\}_{k=1}^n\right) &= %\prod_{k = 1}^K  \left\{ \prod_{i : \xi_i = k} p\left(\boldsymbol{x}^{(i)} \g \btheta_{\xi_i}, \D_{\xi_i}\right) \right\} \\
		\prod_{k=1}^K p\big(\boldsymbol{X}^{(k)}\g\btheta_k,\mathcal{G}_k\big).
		%&= \prod_{k = 1}^ K \left\{\prod_{j = 1}^q \left\{ \prod_{s \in \X_{\pa(j)}} \left\{ \prod_{m \in \Xj} \left\{\thetajpaj\right\} ^{\prescript{}{k}n^{\fa(j)}_{(m,s)}}\right\} \right\} \right\},
	\end{aligned}
\end{equation}
In the equation, $\boldsymbol{X}^{(k)}$ is the $(n_k,q)$ matrix whose rows are the observations $\boldsymbol{x}^{(i)}$ for which $\xi_i=k$, i.e.~that are assigned to cluster $k$; moreover, each term $p\big(\boldsymbol{X}^{(k)}\g\btheta_k,\mathcal{G}_k\big)$ can be written as in \eqref{eq:likelihood:decomposable}.

In the next sections we provide details about the baseline measure $M_0$ of the DP prior. This is structured as $M_0=p(\btheta\g\G)p(\G)$ where in particular $p(\btheta\g\G)$ is a conditional prior on the graph-parameter $\btheta$ given graph $\G$, while $p(\G)$ is a marginal prior on the space of decomposable graphs.
With regard to the DP concentration parameter, we further assign $\alpha \sim \textnormal{Gamma}(c,d)$ following \cite{escobar}.

\subsection{Baseline over graph-parameters}

Let $\G$ be a decomposable UG with set of cliques $\mathcal{C}$ and separators $\mathcal{S}$. Conditionally on $\G$, we assign a prior to $\btheta$, $p(\btheta\g\G)$,
through a Hyper-Dirichlet (HD) distribution \citep{Dawid:Laur:1993}.
Under such prior, $p(\btheta\g\G)$ is strong hyper Markov, meaning that it admits the factorization
\ben
\label{eq:prior:factorization}
p\big(\btheta\g \G\big) \,=\,
\frac{\prod_{C \in \mathcal{C}} p\big(\btheta_C\big)}
{\prod_{S \in \mathcal{S}} p\big(\btheta_S\big)},
\een
where, for each $C\in \mathcal{C}$,  
$p(\btheta_C)$ is the p.d.f.~of a Dirichlet distribution
%$\btheta_C\sim\textnormal{Dir}(\ba^C)$,
%a Dirichlet distribution 
with hyperparameter $\ba^C = \big(a^C_{x_C}, x_C \in \X_C\big)$;
similarly for each $S \in \mathcal{S}$.
Accordingly, we write $\btheta_C\sim\textnormal{Dir}(\ba^C)$ and
\ben
p(\btheta_C) =
\frac{\Gamma\big(\sum_{x_c\in\X_C} a^C_{x_C} \big)}
{\prod_{x_C\in\X_C}\Gamma\big(a^C_{x_C}\big)}
\prod_{x_C\in\X_C}\left\{\theta^C_{x_C}\right\}^{a^C_{x_C}-1}.
\een
The collection $\big\{p(\btheta_C), p(\btheta_S),$ $C \in \mathcal{C}, S \in \mathcal{S}\big\}$
determines a unique hyper Dirichlet prior for $\btheta$ provided that $\big\{p(\btheta_C)$, $C \in \mathcal{C}\big\}$ are \emph{hyperconsistent}, meaning that
\be
%\label{eq:hyperconsistency}
\sum_{x_C \supset x_{C\cap C'}} a^C_{x_C} \,=\, \sum_{x_{C'} \supset x_{C\cap C'}} a^{C'}_{x_{C'}}
\ee
for any $x_{C\cap C'} \in \X_{C\cap C'}$ and any two cliques $C$ and $C'$ such that $C \cap C' \neq \O$; see in particular \citet[p.~1304]{Dawid:Laur:1993}.
Hyperconsistency is therefore fully reflected in the choice of the Dirichlet hyperparameters; in particular it is guaranteed by any value $a^C_{x_C}=a/|\X_C|$ with $a>0$.
%since the number of terms in the left and right-side sum of \eqref{eq:hyperconsistency} is equal to $|\X_C|$ and $|\X_{C'}|$ respectively.
In the following we consider the default choice $a=1$; % so that $a^C_{x_C}=1/|\X_C|$;
see also \citet{Castelo:Perlman:2004}.

\subsection{Baseline over graphs}

Let $\bA^\G$ be the $(q,q)$ adjacency matrix of graph $\G$.
To induce a prior on $\G$ we start by assigning
\be
\bA^\G_{u,v}\g\pi &\stackrel{\textnormal{iid}}\sim& \textnormal{Bern}(\pi), \quad u>v, \\
\pi &\sim& \textnormal{Beta}(a_{\G},b_{\G}),
\ee
that is, conditionally on $\pi\in(0,1)$, a collection of independent Bernoulli priors on the 0-1 elements of the lower-triangular part of $\bA^\G$, and a Beta distribution on the (prior) probability of edge inclusion $\pi$.
Accordingly, the induced marginal prior on $\G$ is
\ben
\label{eq:prior:graph}
p(\G) \propto \frac{\Gamma\left(a_{\G} + |\G|\right)\Gamma\left(b_{\G}+\frac{q(q-1)}{2}-|\G|\right)}{\Gamma\left(\frac{q(q-1)}{2}+a_{\G}+b_{\G}\right)}\cdot \frac{\Gamma\left(a_{\G}+b_{\G}\right)}{\Gamma\left(a_{\G}\right)\Gamma\left(b_{\G}\right)}, \quad \G\in\mathcal{P}_\G,
\een
where $|\G|$ is the number of edges in the graph $\G$, and $\mathcal{P}_\G$ is the set of all decomposable UGs with $q$ nodes.
Hyperparameters $a_\G$ and $b_\G$ can be tuned in order to control the expected prior probability of edge inclusion in the graph; in particular, the default choice $a_\G=b_\G=1$ implies a uniform prior on $\pi$, while any choice $a_\G<b_\G$, with $a_\G,b_\G > 0$, will favor graph sparsity.

\vspace{0.5cm}

\section{Posterior inference}
\label{sec:Computational:details}

We conduct posterior inference of our model through a partially collapsed sampler which targets the marginal posterior of cluster indicators, graphs and concentration parameter
$p\big(\{\xi_i\}_{i = 1}^n, \{\mathcal{G}_k\}_{k =1}^K,$ $\alpha, K \g \bX \big)$,
where $K$ is the number of non-empty clusters, i.e.~unique values among the $\xi_i$'s; see also \citet{Rodriguez:et:al:2009}. %\citep[Algorithm 4]{Neal:2000}.
Our algorithm maintains a Gibbs-sampling structure, involving, for a number of iterations, the following three steps:
\begin{enumerate}	
	\item Sample the cluster indicators $\{\xi\}_{i = 1}^n$ from their full conditional $p(\{\xi_i\}_{i = 1}^n \g \textnormal{rest})$;
	\item Sample the graphs $\{\mathcal{G}_k\}_{k=1}^K$ from their full conditional $p(\{\mathcal{G}_k\}_{k =1}^K \g \textnormal{rest})$;
	\item Sample the DP concentration parameter $\alpha$ from its full conditional $p(\alpha\g \textnormal{rest})$.
\end{enumerate}

\vspace{0.5cm}

The full conditional of cluster indicator $\xi_i$ is
	\begin{equation}
		\begin{aligned}
			p\big(\xi_i = k \g \{\boldsymbol{x}^{(l)} : l \neq i, \xi_l = k\}, \mathcal{G}_k\big) \hspace{5cm} \\ \propto
			\begin{cases}
				\,n^{-i}_k \, p\big(\boldsymbol{x}^{(i)} \g  \{\boldsymbol{x}^{(l)} : l \neq i, \xi_l = k\}, \mathcal{G}_k\big) & k = 1,\dots,K \\
				\,\alpha \, p\big(\boldsymbol{x}^{(i)} \g \mathcal{G}_{k}\big) & k = K + 1,
			\end{cases}
		\end{aligned}
	\end{equation}
where $n^{-i}_k$ is the number of observations currently associated with cluster $k$, excluding individual $i$. The first case of the equation deals with non-empty clusters. In particular, the (conditional) probability of assigning individual $i$ to cluster $k$ is proportional to the posterior predictive probability of $\boldsymbol{x}^{(i)}$ conditionally on all (other) observations currently assigned to cluster $k$. The second case, instead, deals with the assignment of individual $i$ to a new, thus empty, cluster $K+1$. $p\big(\xi_i = k \g \cdot\big)$ is proportional to the prior predictive probability based on observation $\boldsymbol{x}^{(i)}$ only. Notice that, implicitly, the same step provides an update of $K$, i.e.~the number of non-empty clusters. The following two propositions provide the closed-form expressions of the posterior and prior predictive distributions above. Proofs of the results are reported in the Supplementary Material.

\begin{prop}[Posterior predictive distribution - non-empty cluster]
The posterior predictive distribution of $\bx^{(i)}$, given all the other observations currently assigned to cluster $k$,  $\{\boldsymbol{x}^{(l)} : l \neq i, \xi_l = k\}$, and the graph $\mathcal{G}_k$ having set of cliques and separators $\mathcal{C}_k$ and $\mathcal{S}_k$ respectively is such that: 
\ben
\begin{aligned}
\label{eq:predictive:non:empty}
p\big(\bx^{(i)} \g \{\boldsymbol{x}^{(l)} : l \neq i, \xi_l = k\}, \G_k\big)
&=
\big\{
a + n_k - \mathbbm{1}\{\xi_i = k\}
\big\}^{|\mathcal{S}_k| - |\mathcal{C}_k|}
\\
&\,\, \cdot \,\,\frac{
\prod_{C \in \mathcal{C}_k}
\Big(
a_{\bx^{(i)}_C}^C
+ \prescript{}{k} n_{\bx^{(i)}_C}^{C}
- \mathbbm{1}\{\xi_i = k\}
\Big)
}{
\prod_{S \in \mathcal{S}_k}
\Big(
a_{\bx^{(i)}_S}^S
+ \prescript{}{k} n_{\bx^{(i)}_S}^{S}
- \mathbbm{1}\{\xi_i = k\}
\Big)
},
\end{aligned}
\een 
where, for a generic $S \subseteq V$,  $a_{\bx^{(i)}_S}^S$ denotes the element of $\boldsymbol{a}^S$ corresponding to the observed configuration $\bx^{(i)}_S$, and $ \prescript{}{k} n_{\bx^{(i)}_S}^{S} = \sum_{l: \xi_l = k} \mathbbm{1} \{ \bx^{(l)}_S = \bx^{(i)}_S\}$.
\begin{proof}
	See Supplementary Material (Section 2).
\end{proof}
\end{prop}

\begin{prop}[Prior predictive distribution - empty cluster]
The prior predictive distribution of $\bx^{(i)}$ given the graph $\mathcal{G}_k$ having set of cliques and separators $\mathcal{C}_k$ and $\mathcal{S}_k$ respectively is such that:
\ben
\label{eq:marginal:empty}
p\big(\bx^{(i)} \g \G_k\big)
\,=\,
\big\{a\big\}^{|\mathcal{S}_k|-|\mathcal{C}_k|}
\cdot
\frac
{\prod_{C \in \mathcal{C}_k}a_{\bx^{(i)}_C}^C}
{\prod_{S \in \mathcal{S}_k}a_{\bx^{(i)}_S}^S}, 
\een
where for a generic $S \subseteq V$, $a_{\bx^{(i)}_S}^S$ denotes the element of $\boldsymbol{a}^S$ corresponding to the observed configuration $\bx^{(i)}_S$.
\begin{proof}
	See Supplementary Material (Section 2).
\end{proof}
\end{prop}

\vspace{0.5cm}

The full conditional of graphs  $\{\mathcal{G}_k\}_{k=1}^K$ does not admit a closed-form expression instead. Accordingly, we can sample from it through a Metropolis Hastings scheme as follows. For each non-empty cluster $k$, we first propose a candidate graph from a suitable proposal $q(\widetilde{\mathcal{G}}_k\g\mathcal{G}_k)$ conditionally on the current $\mathcal{G}_k$. Then, $\widetilde{\mathcal{G}}_k$ is accepted with probability
\begin{equation}
		\alpha_{\widetilde{\mathcal{G}}_k}=\min\left\{1;
		\frac{m_{}(\boldsymbol{X}^{(k)}\g \widetilde{\mathcal{G}}_k)}
		{m_{}(\boldsymbol{X}^{(k)}\g \mathcal{G}_k)}
		\cdot\frac{p(\widetilde{\mathcal{G}}_k)}{p(\mathcal{G}_k)}
		\cdot\frac{q(\mathcal{G}_k\g\widetilde{\mathcal{G}}_k)}{q(\widetilde{\mathcal{G}}_k\g\mathcal{G}_k)}\right\},
\end{equation}
where in particular $m_{}(\boldsymbol{X}^{(k)}\g \mathcal{G}_k)$ is the marginal likelihood of graph $\G_k$ based on the data $\boldsymbol{X}^{(k)}$.
Full details relative to the proposal distribution and the expression of the marginal likelihood are provided in the Supplementary Material.

\vspace{0.5cm}

The full conditional distribution of $\alpha$ reduces to $p(\alpha\g K)\propto p(K\g \alpha)p(\alpha)$, where in particular
$$
p(K\g\alpha)\propto c_n(K)\alpha^K \frac{\Gamma(\alpha)}{\Gamma(\alpha+n)}
$$
is the prior on the number of clusters induced by the DP prior,
and $c_n(K)$ is a normalizing constant not depending on $\alpha$.
Sampling from $p(\alpha\g K)$ can be done by augmenting the distribution through an auxiliary variable $\eta \sim \textnormal{Beta}(1, \alpha)$. It can be shown \citep{escobar}
that under the prior $\alpha \sim \textnormal{Gamma}(c,d)$ the full conditional of $\alpha \g K, \eta$ corresponds to a mixture of Gamma distributions, specifically
$$
p(\alpha \g \eta, K) = g \cdot \textnormal{dGamma}(\alpha; c+K, d-log \eta) + (1-g) \cdot \textnormal{dGamma}(\alpha;c+K-1, d-log \eta),
$$
where $g/(1-g) =  (c + K -1) / n(d - log \eta)$ and $\textnormal{dGamma}(\alpha;a,b)$ denotes the p.d.f.~of a Gamma distribution with shape and rate parameters $a$ and $b$ respectively.

\subsection{Posterior summaries}

Our MCMC scheme approximates the posterior over cluster indicators and graphs by providing a sample of $S$ draws from this target distribution. Specifically, let
$K^{(s)}$ be the number of clusters at MCMC iteration $s$, while
$\xi_i^{(s)}$, $i=1,\dots,n$, and $\G_k^{(s)}$, $k=1,\dots,K^{(s)}$, the corresponding sampled values of cluster indicators and graphs.
We can provide posterior summaries about the learned clustering structure by
estimating an $(n,n)$ posterior similarity matrix $\boldsymbol{\widehat S}$. The $(i,i')$-element of the matrix corresponds to the estimated posterior probability of subjects $i$ and $i'$ belonging to the same cluster, and is given by
\begin{equation} \label{eq:post:prob:same:cluster}
	\widehat{p}(\xi_i = \xi_{i'} \g \bX) = \frac{1}{S} \sum_{s = 1}^S \mathbbm{1}\left\{\xi_i^{(s)} = \xi_{i'}^{(s)}\right\}.
\end{equation}
The ensuing collection of probabilities allows to quantify the uncertainty regarding an existing clustering structure in the data. Moreover, a point estimate of the clustering, $\widehat{\boldsymbol{c}}$, can be recovered as the partition minimizing an expected loss function, such as the Variation of Information (VI); see in particular \cite{wade:2018}.

From the same MCMC output, we can also recover for each individual $i$ a $(q,q)$ matrix collecting estimates of the Posterior Probabilities of edge Inclusion (PPIs). For a given edge $u - v, u \neq v$ and subject $i$, its estimated PPI is
\begin{equation} \label{eq:post:prob:edge:inclusion}
	\widehat{p}_i (u - v \g \bX) = \frac{1}{S} \sum_{s = 1}^S \mathbbm{1}\left\{u - v \in \G^{(s)}_{\xi_i^{(s)}}\right\},
\end{equation}
which corresponds to the proportion of graphs assigned to individual $i$ in the chain that contain the edge $u - v$.
If required, a subject-specific graph estimate $\widehat{\G}_i$ can be finally obtained by including those edges for which $\widehat{p}_i (u - v \g \bX) > z$, with $z \in (0,1)$, e.g $z = 0.5$.

%%va riscritto tutto il blocco delle simulazioni 
\section{Simulation study}
\label{sec:simulation:study}

In this section we evaluate the clustering performance of our method through simulations.
%\citep{goodman,linzer}
In addition, we include a comparison with the following benchmark methods:
\begin{itemize}
	\item \textit{LCA}: a Latent Class Analysis model for polytomous variables \citep{linzer};
	\item \textit{Bayes LCA}: a Bayesian-model formulation of LCA \citep{White:Murphy:BayesLCA};
	\item \textit{MFM LCA}: a two-layer Mixture of Finite Mixtures Latent Class Analysis model \citep{Malsiner:et:al:2024}.
\end{itemize}

We fix the number of clusters as $K=2$, corresponding to the true graphs $\G_1$ and $\G_2$. We instead range the number of variables $q\in\{10,20\}$, and consider three distinct scenarios. Scenario 0 assumes independence between variables, equivalently $\G_1$ and $\G_2$ having no edges, while Scenario 1 and Scenario 2 are characterized by dependencies between variables. For both, we first generate randomly $\G_1$ as a sparse graph with $q=20$ nodes and number of edges equal to $20$; graph $\G_2$ is obtained by applying a number $M$ of random consecutive local moves (edge additions or deletions) to $\G_1$; we fix $M=10$ in Scenario 1 and $M=20$ in Scenario 2, implying strong and moderate degrees of similarity between the two graphs respectively.
For each $k \in\{1,2\}$, $n_k = 200$ categorical observations are generated (independently across $k$) by discretization of Gaussian data as follows. We first generate i.i.d.~multivariate observations $\{\boldsymbol{y}^{(i)}, i=1,\dots,n_k\}$ from a zero-mean Gaussian UG model, whose precision (inverse-covariance) matrix reflects the conditional independencies of $\G_k$. After standardization of each variable, we set $\boldsymbol{x}^{(i)}_j = 0$ if $\boldsymbol{y}^{(i)}_j < z_{\alpha_j}$, and $1$ otherwise, where $z_{\alpha_j}$ denotes the quantile of order $\alpha_j$ of a standard Normal distribution. Values of the $\alpha_j$ coefficients determine the empirical marginal probabilities of the resulting binary categorical data. These probabilities are kept fixed across the three simulation scenarios, so that differences in the clustering performance can be attributed to changes in the dependence structure across clusters. %, rather than differences in the marginal distributions.
Under each scenario we consider $N=20$ independent simulations corresponding to $20$ simulated datasets with $q=20$ categorical variables.
The analysis for the $q=10$ setting is conducted under the three scenarios above by considering the first $10$ variables in each dataset. %and the corresponding sub-graphs of $\G_1$ and $\G_2$ as benchmarks. Non occorrono i grafi come benchmark perché non valutiamo structure learning.

%inserisci table

%For both methods, we input the number of clusters as $K=2$.
%Performances are assessed by comparing the true partition $\boldsymbol{c}$ with the estimated partitions $\widehat{\boldsymbol{c}}$ based on the Variation of Information (VI). Lower values of the metric correspond to better performances. In addition, we expect scenario with $\alpha = 0.1$ to be characterized by overall better performances than $\alpha = 0.4$, because in the latter the difference between the two clusters is mainly due to the dependency structure among the variables. Results are summarized in the boxplots of Figure \ref{fig:clustering_sim}.

\begin{figure}[H]
	\centering
	\setlength{\tabcolsep}{5pt} % space between columns
	\begin{tabular}{cc}
		\vspace{0.4cm}
		% ---- Row 1 ----
		\raisebox{-5mm}{\rotatebox{90}{Scenario 0}} &
		\begin{minipage}{0.8\textwidth}
			\centering
			\includegraphics[height=0.28\textheight]{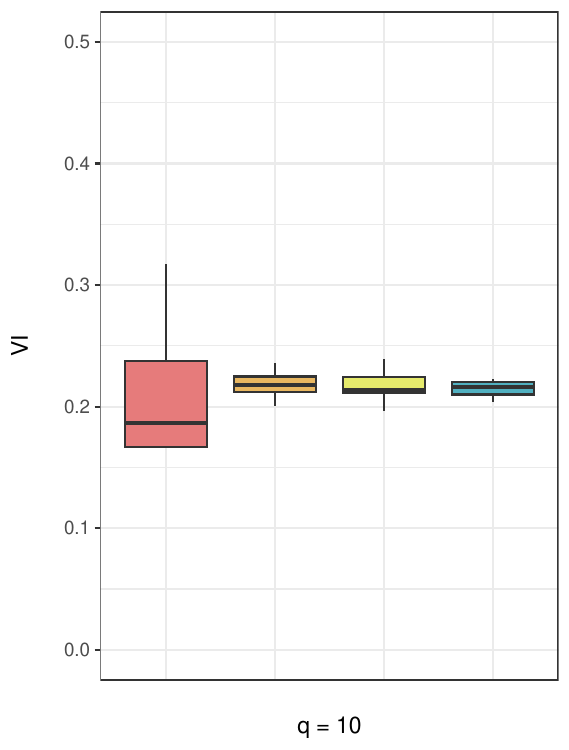}
			\includegraphics[height=0.28\textheight]{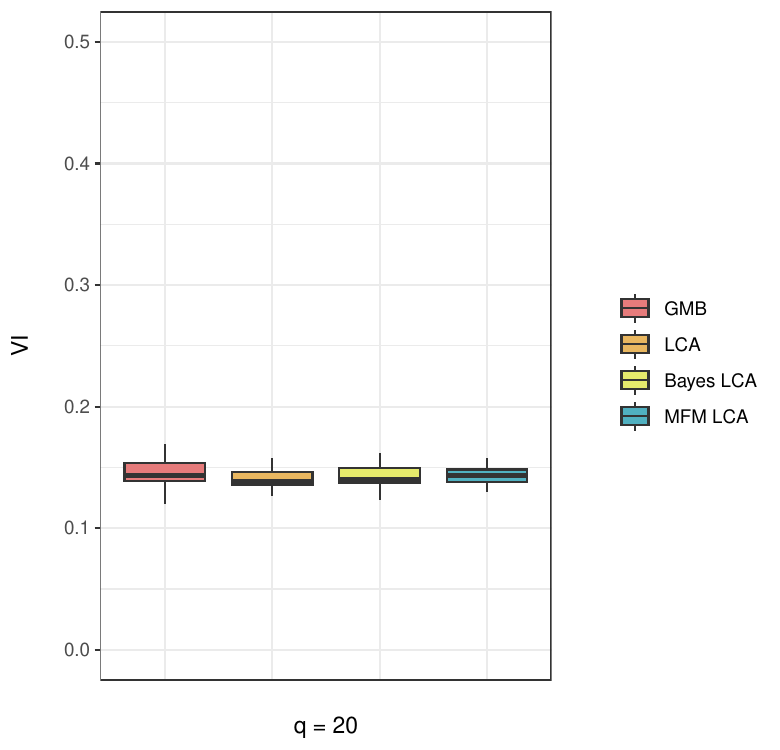}
		\end{minipage} \\ \vspace{0.4cm}
		
		% ---- Row 2 ----
		\raisebox{-5mm}{\rotatebox{90}{Scenario 1}} &
		\begin{minipage}{0.8\textwidth}
			\centering
			\includegraphics[height=0.28\textheight]{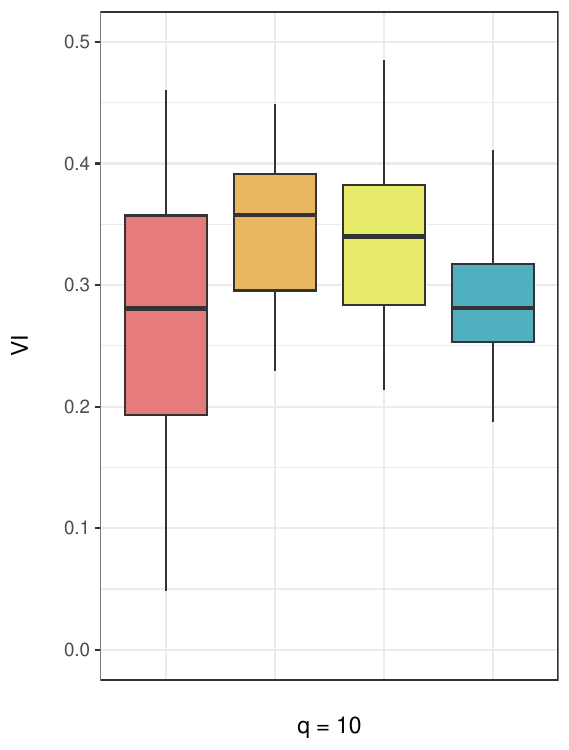}
			\includegraphics[height=0.28\textheight]{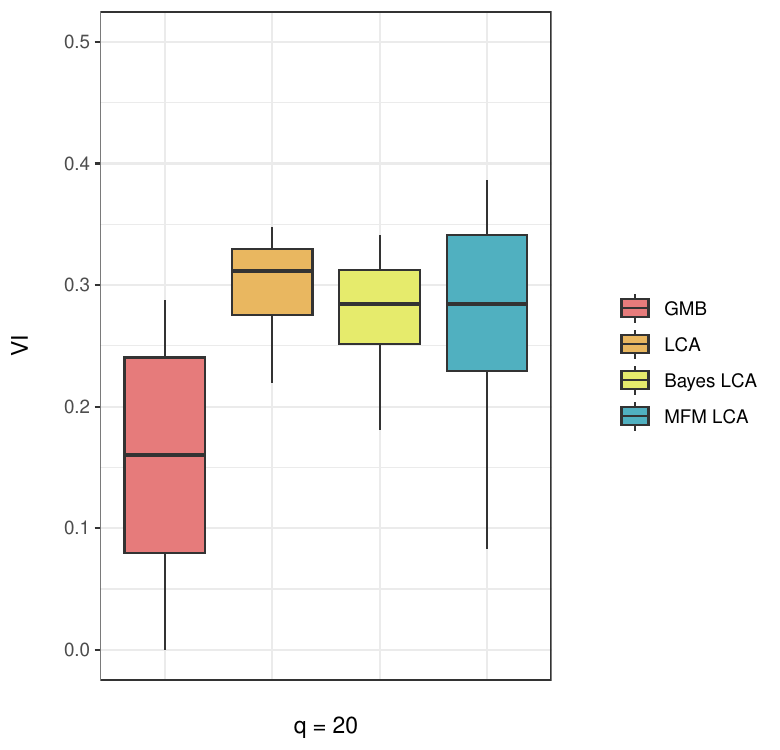}
		\end{minipage} \\ \vspace{0.2cm}
		
		% ---- Row 3 ----
		\raisebox{-5mm}{\rotatebox{90}{Scenario 2}} &
		\begin{minipage}{0.8\textwidth}
			\centering
			\includegraphics[height=0.28\textheight]{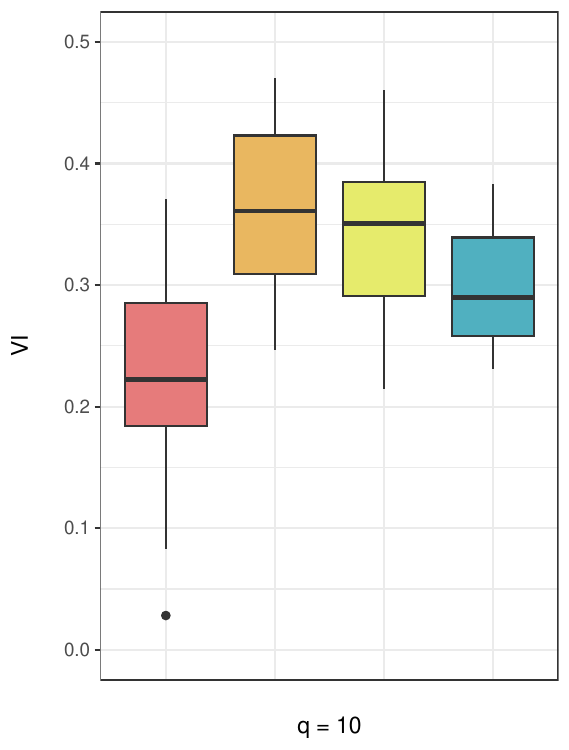}
			\includegraphics[height=0.28\textheight]{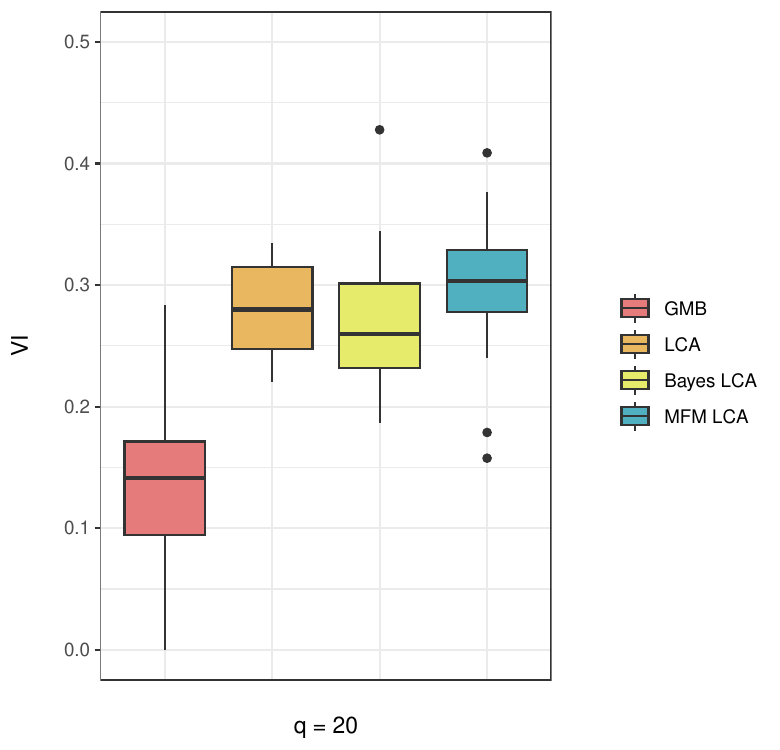}
		\end{minipage}
	\end{tabular}
	\caption{Simulations. Distribution of the Variation of Information (VI) between true and estimated clusterings for values of $q\in\{10,20\}$ and each method under comparison. The three scenarios reflect different degrees of dependence and similarity between the two cluster-specific graphs: Scenario 0 is characterized by equal graphs with no edges; Scenario 1 and Scenario 2 by graphical structures having distance $M=10$ and $M=20$ respectively.}
	\label{fig:sims}
\end{figure}

With regard to our Graphical Model-Based clustering method (\textit{GMB}),
%we set $a=1$ in the Hyper Dirichlet prior on $\btheta$, \red l'abbiamo già data come scelta diciamo di default nella sezione 3; forse si può rimuovere da qui, visto che non fa neanche riferimento ad una equazione; nella (7) non compare ancora $a$ ma viene definito nel testo dopo \black
we fix $a_{\G}=1,b_{\G}=3$ in the prior on graphs \eqref{eq:prior:graph}, a choice which can favor sparsity in the dependence structure, and $c=3$, $d=1$ in the Gamma prior on the DP concentration parameter $\alpha$. 
Moving to the benchmark methods, both \textit{LCA} and \textit{Bayes LCA} consider a fixed number of clusters $K$; inference on $K$ is thus tackled as a model selection problem by running each method for a grid of values of $K\in\{2,3,4,5\}$, and then selecting the model with the best fitting according to the Bayesian Information Criterion (BIC).
%% commenti trade off kernel density estimation e numero di cluster --> aspettiamoci che il numero di cluster stimato sia piu grande
The \textit{MFM LCA} method is instead implemented following the guidelines provided by \citet{Malsiner:et:al:2024}. Specifically, the prior on the number of mixture components is set as a Beta-Negative Binomial distribution with parameters $(1,3,4)$.
%Additionally, the joint tuning of \textit{cluster cohesion parameter} $a_{\mu}$, \textit{class cohesion parameters} $c_{\phi}$ and the number of classes in the inner layer $L$ has been shown to have an impact on the clustering performance \red questo mostrato da loro o da noi in alcune prove? \black
In addition, following the results in their simulation studies, we set the cluster cohesion parameter $a_{\mu} = 20$, the class cohesion parameter $c_{\phi} = 30$, and the number of classes in the inner layer $L = 4$.
While \textit{LCA} directly estimates a clustering partition, the other Bayesian methods provide a posterior similarity matrix; see also our Equation \eqref{eq:post:prob:same:cluster}; from the latter, a point estimate of the clustering is obtained as the partition minimizing the expected Variation of Information (VI); see again \citet{wade:2018}.
The VI metric is also employed to assess the clustering performances of the various methods, namely as a measure of distance between the true and any of the estimated clustering partitions, $\boldsymbol{c}$ and $\widehat{\boldsymbol{c}}$ respectively.
Results, under each of the three simulation scenarios, and for values of $q\in\{10,20\}$ are summarized in the box-plots of Figure \ref{fig:sims}.

From the results, it appears how the clustering performance of all methods overall improves as the number of variables $q$ increases. In Scenario 0, which is characterized by independence between variables, the various methods tend to perform similarly. In the next two scenarios instead, our methodology clearly outperforms the benchmarks. This is due to the fact that our graphical-model based method explicitly accounts for possible differences between clusters in the existing dependence structure across variables. Among the competing approaches, \emph{MFM LCA} generally performs better than the other methods, with comparable results only in Scenario 2 for $q = 20$. This behavior could likely be improved by tuning the prior hyperparameters, for instance by increasing the number of inner components $L$, which could give a higher flexibility in capturing the within-cluster associations among variables.

\section{Real data analyses}
\label{sec:Application}
The following section presents real-world applications of our new methodology. The first is based on a biological dataset of allelic variants in a study on cystic fibrosis. The second is instead based on voting opinions, and aim at identifying possible latent subgroups within republican and democratic parties. %In both applications, we expect that accounting for dependencies among variables may improve the clustering performance, since the assumption of conditional independence may be too restrictive, as discussed later.
In both cases, the analyzed data reveal the existence of explicit dependencies between variables, which can therefore characterize possible clustering structures.

\subsection{Cystic fibrosis data}
\label{sec:Application:fibrosis}
A primary goal in genomic case-control studies is to identify associations between genetic variants and disease-related traits. To this end, multiple genetic variants located on a chromosome, known as haplotypes, are measured on both individuals that are affected by a disease, as well as on a control group, that is disease-free patients. These haplotypes tend to be inherited together, a phenomenon referred to as Linkage Disequilibrium (LD), which reflects the non-random association of alleles at closely-located loci or markers \citep{Schaid:et:al:2018}. It is well known that LD patterns can differ between case and control populations, and may also exhibit additional sources of heterogeneity, for instance due to differences in ancestry or ethnicity.
As a consequence, multivariate statistical methods applied to such data should account for possible associations among allelic frequencies at different loci.
In this work, we consider genomic data from a medical study on cystic fibrosis. The analysed dataset, originally presented in \citet{Kerem:et:al:89}, includes $n = 160$ individuals, of whom $82$ belong to the control group, and measurements collected at $q = 23$ loci. 
%Each locus takes one of two allelic states, coded as $0$ or $1$, where $0$ and $1$ are arbitrary categorical labels indicating the two possible alleles at that locus. Instances of missing data are present as well and are handled as an additional categorical level. \black
Information at each locus is recorded as a categorical variable with levels $0$ and $1$ indicating two possible alleles, and a third level in case of missing information.
The same dataset has previously been analyzed, among others, by \citet{Sabatti:et:al:2001} for target discovery purposes, specifically to identify loci that are potentially linked to disease-causing mutations and in particular those affected by changes in the allelic frequencies between cases and controls.
%\gray see also [REF noi].
%\red Non ci citerei per ora, visto cmq l'overlap tra i due lavori \black
%%%%%%%%%%%%%
%% Citare ns lavoro quando ci sarà arXiv
%%%%%%%%%%%%%
We consider a different perspective, and apply our methodology to cluster individuals on the basis of the available genetic information. These clusters are expected to reflect similarities in the genetic variants that characterize the true group memberships, namely cases and controls. At the same time, the analysis may reveal additional sources of heterogeneity in the underlying LD structure.

\begin{figure}[H]
	\centering
	\includegraphics[width=.65\linewidth]{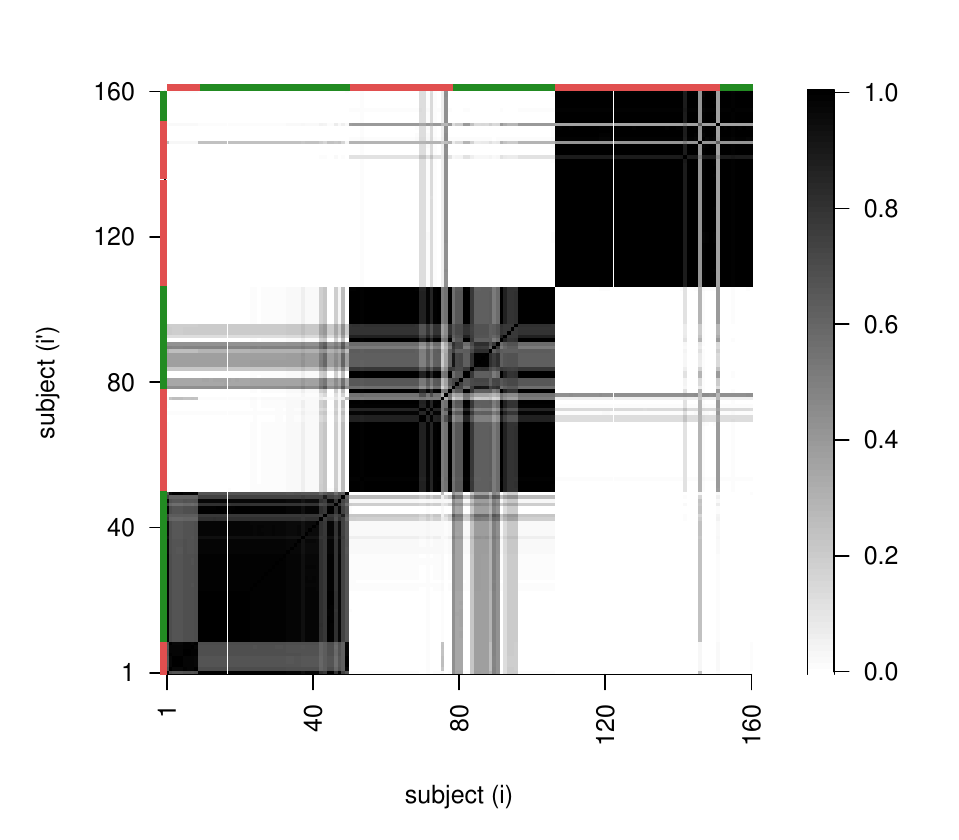} 
	\vspace*{-3mm}
	\caption{Cystic fibrosis data. Estimated posterior similarity matrix, with subjects arranged according to the estimated clustering partition, and colors representing their disease status, red and green for cases and controls respectively.}
	\label{fig:fibrosis:psm}
\end{figure}

We set prior hyperparameters as $a=1$ in the Hyper Dirichlet prior, $a_{\G}=b_{\G}=1$ in the prior on graphs, resulting in a uniform prior probability of edge inclusion, and $c=3$, $d=1$ in the Gamma prior on the DP concentration parameter.
Our algorithm is then implemented for $S=120000$ iterations after a burn in period of $20000$ draws, and some pilot runs that were used to assess the MCMC convergence by monitoring the agreement between multiple chains.
Figure \ref{fig:fibrosis:psm} provides a graphical representation of the estimated posterior similarity matrix, with subjects arranged according to a point estimate of the clustering partition obtained \emph{via} the expected VI criterion \citep{wade:2018}. In particular, the $(i,i')$-cell of the matrix corresponds to the estimated probability of subjects $i$ and $i'$ belonging to the same cluster; see also Equation \eqref{eq:post:prob:same:cluster}. Colors at the borders of the matrix instead labels the $n=160$ patients according to their disease status, namely case (in red) or control (in green).

\begin{figure}[H]
	\centering
	\includegraphics[width=.8\linewidth]{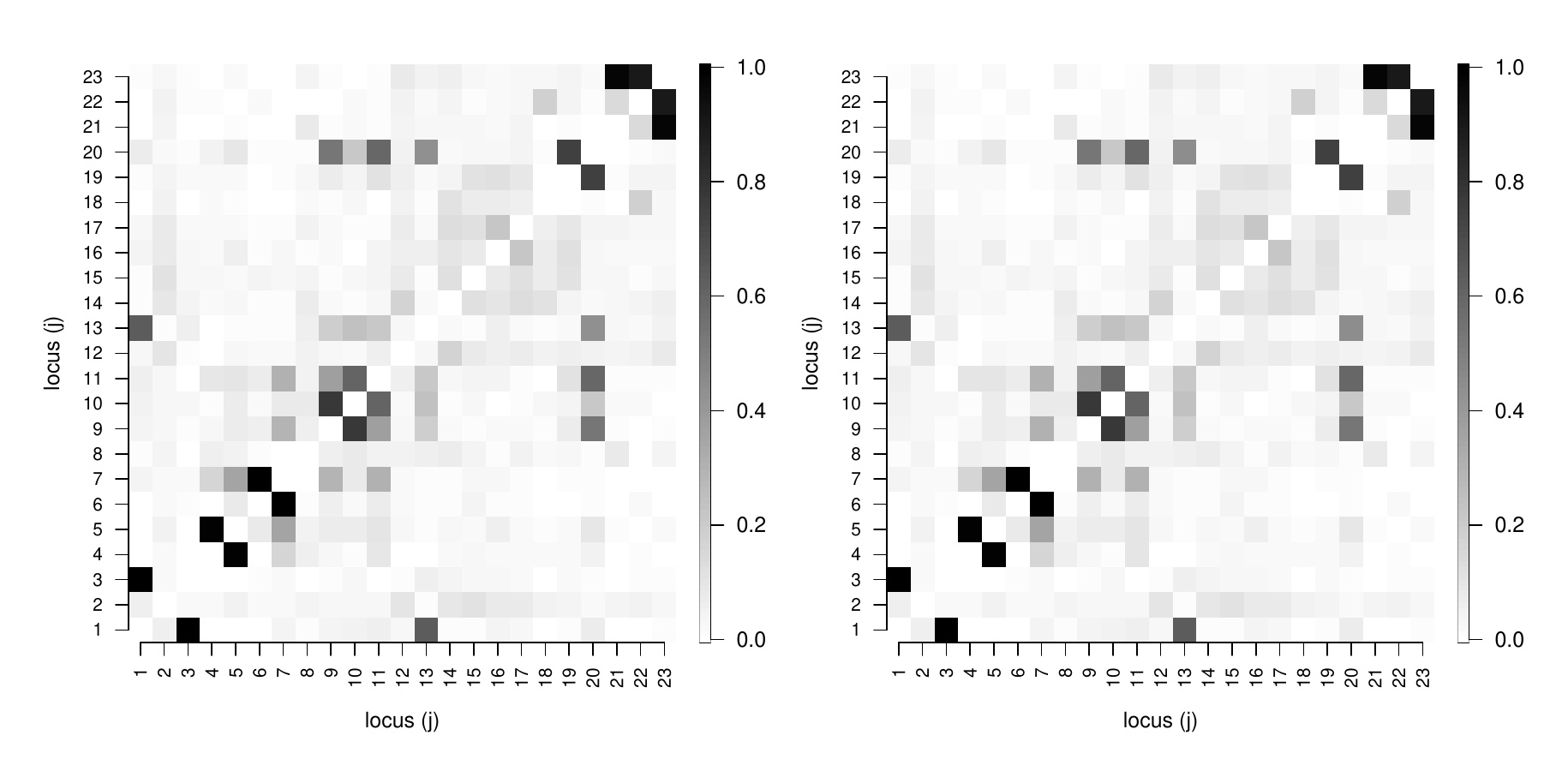} \\
	\includegraphics[width=.8\linewidth]{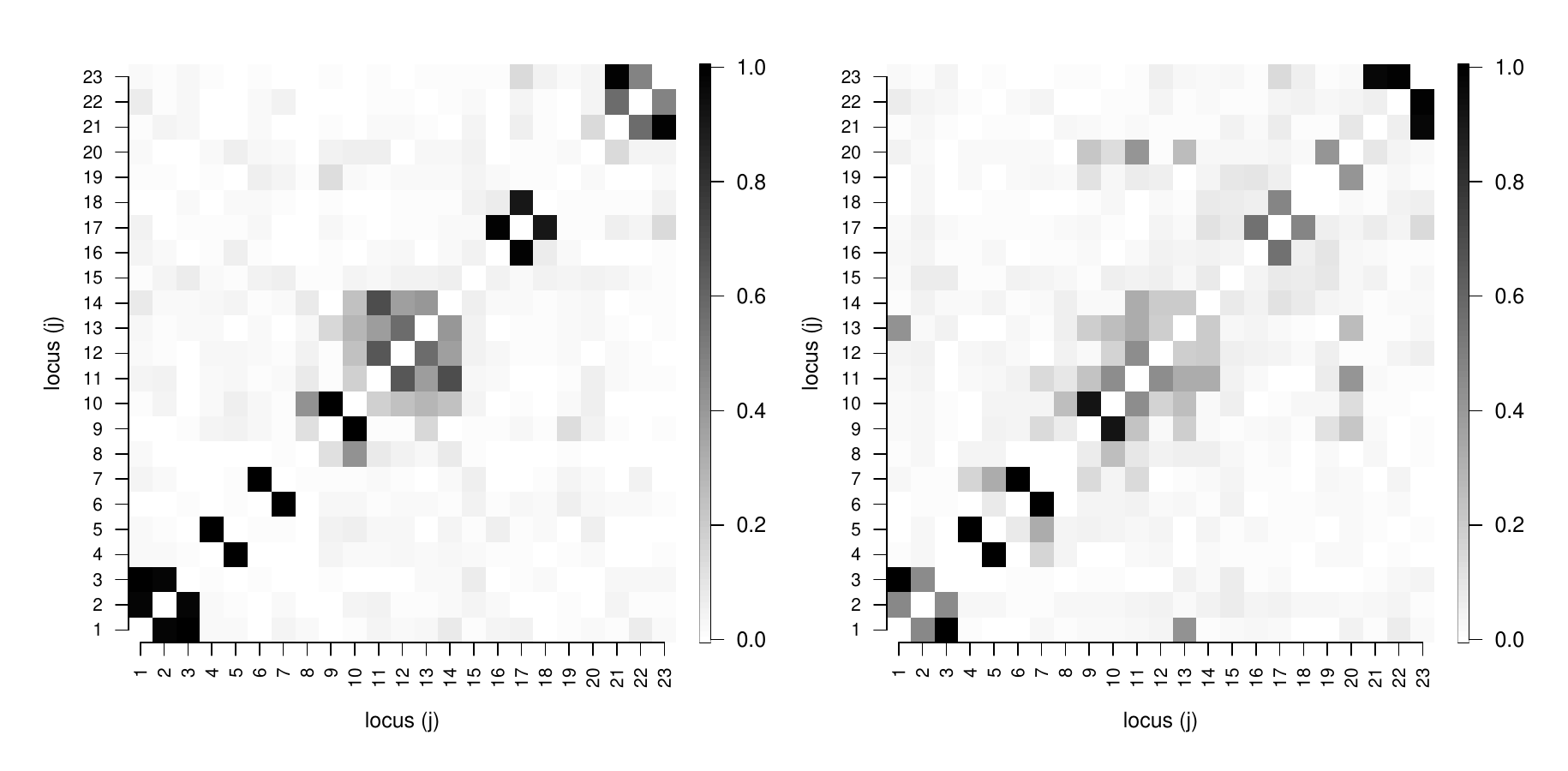} \\
	\includegraphics[width=.8\linewidth]{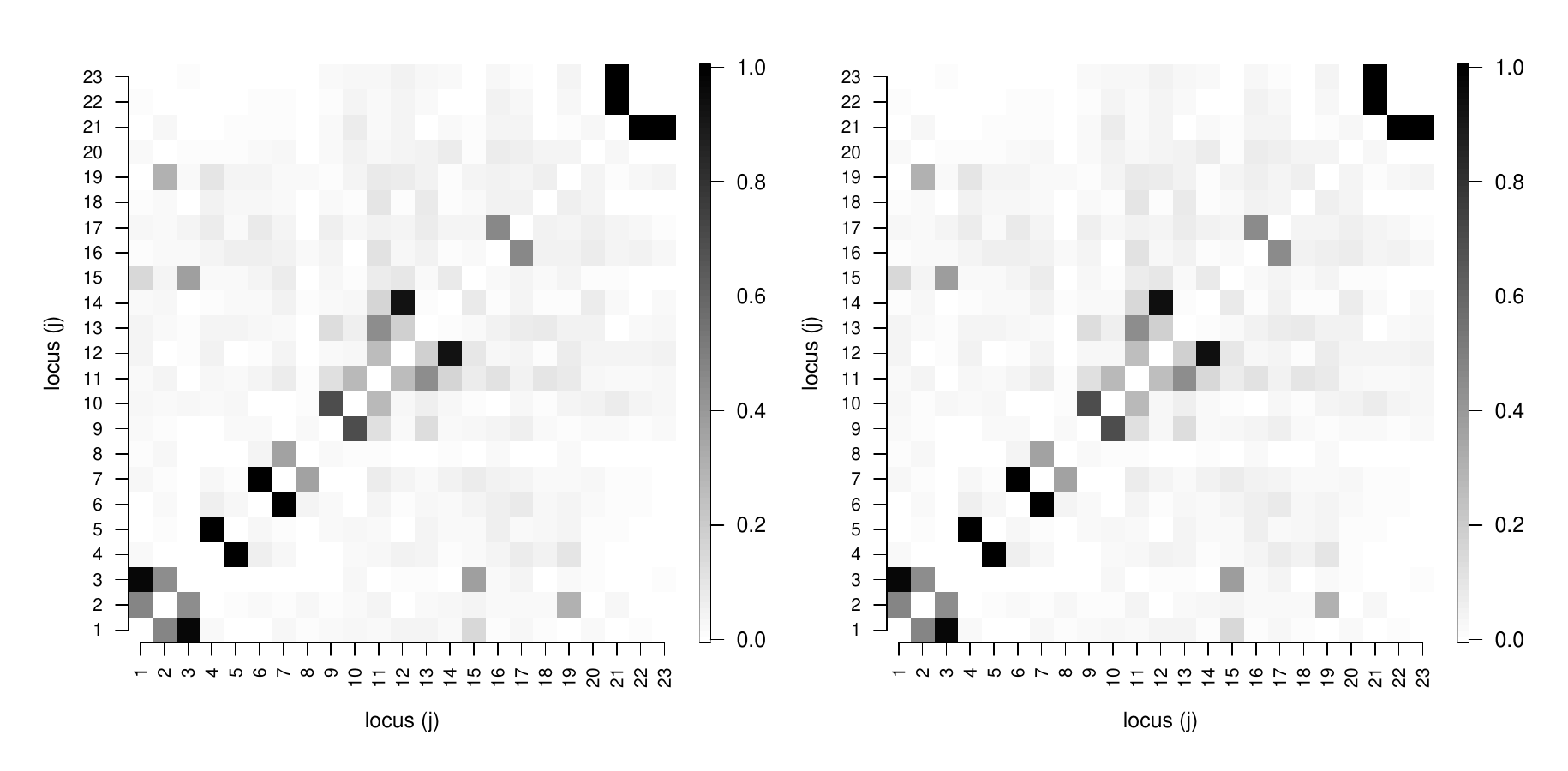}
	\caption{Cystic fibrosis data. Estimated posterior probabilities of edge inclusion for two randomly chosen individuals within each of the three estimated clusters (one row for each cluster).}
	\label{fig:ppi}
\end{figure}

\begin{figure}[H]
    \centering
    \quad \quad \quad \quad \includegraphics[width=0.38\textwidth]{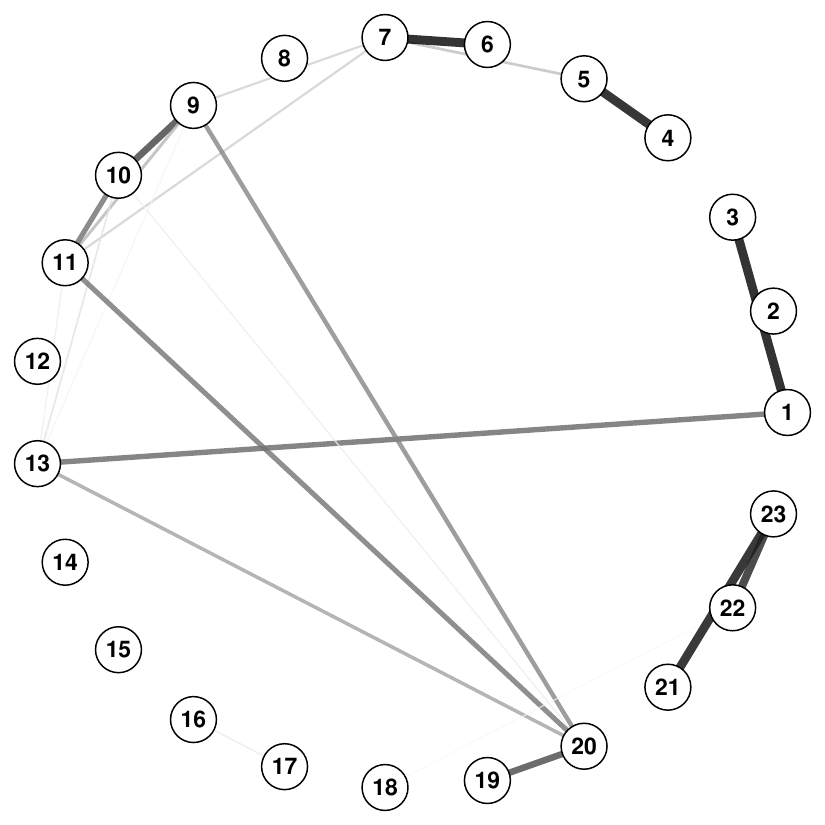}
    \hfill
    \includegraphics[width=0.38\textwidth]{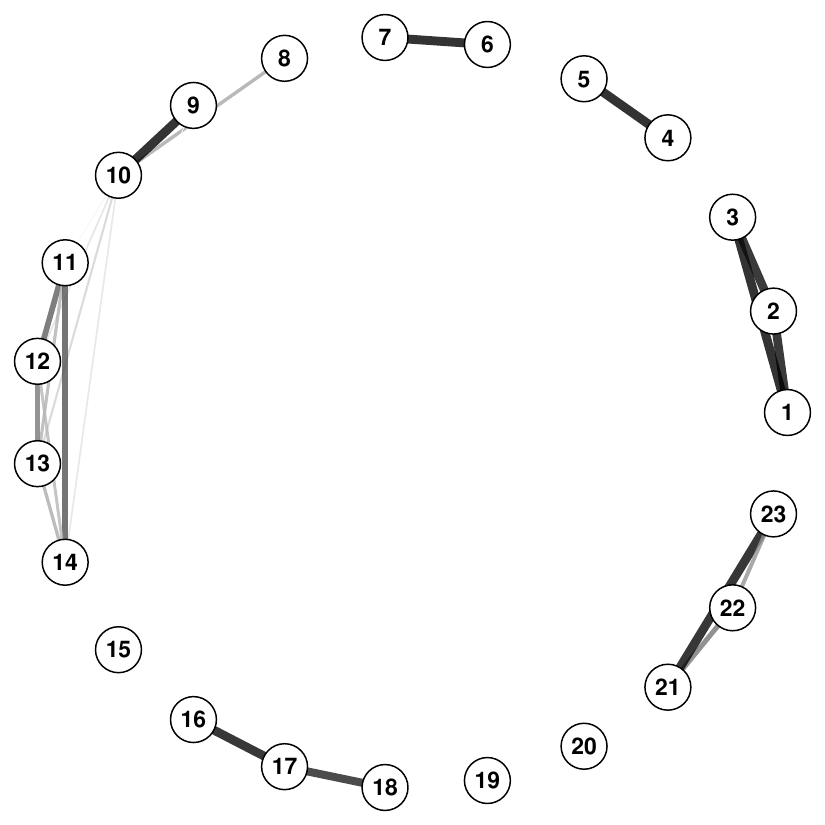} \quad \quad \quad \quad \\
    %\hfill
    \includegraphics[width=0.38\textwidth]{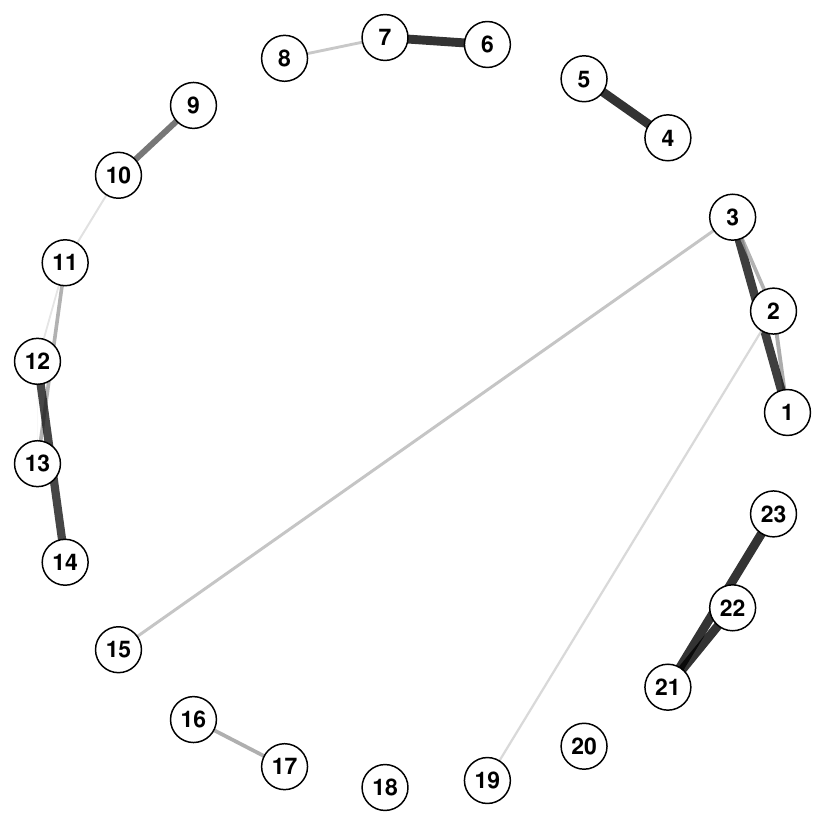}
    \\
    \hspace{0.2cm}
    \caption{Cystic fibrosis data. Graph representation of estimated subject-specific posterior probabilities of edge inclusion representing the dependence structure between loci. Each graph refers to a subject randomly selected from one of the three estimated clusters.}
    \label{fig:graph}
\end{figure}

A three-clustering structure is clearly visible from Figure \ref{fig:fibrosis:psm}. A first cluster mostly consists of control patients, while in a second one most patients are affected by cystic fibrosis. This suggests that the method is able to identify disease-specific heterogeneity in LD patterns. In contrast, a third cluster contains cases and controls in approximately equal proportion, suggesting that the heterogeneity in LD for this group is likely driven by other latent characteristics. %Similarly to what was done in the simulations, we obtain a point estimate of the clustering structure by means of VI. \newline 
Figure \ref{fig:ppi} reports the estimated posterior probabilities of edge inclusion for two individuals randomly selected from each of the three estimated clusters. Interestingly, individuals within the same cluster (appearing in the same row of the plot) display similar dependence patterns, while substantial differences emerge across clusters. Despite these differences, the three clusters are characterized by higher probabilities of inclusion for edges connecting loci that are closely located. This is consistent with the biological knowledge that adjacent loci tend to exhibit a stronger association.
These patterns are clearly visible from Figure \ref{fig:graph} too, which provides a graph-based representation of the above posterior probabilities of edge inclusion for one randomly-chosen subject from each estimated cluster.

\subsection{Voting records}
\label{sec:Application:voting}
We consider the popular voting records data collected from the 1984 United States Congress. The dataset, publicly available at \url{https://archive.ics.uci.edu/}, includes individual votes for each of the $n = 434$ U.S. House of Representatives Congressmen, corresponding to $n_D = 267$ democrats and $n_R = 167$ republicans, relative to $q=16$ key votes identified by the Congressional Quarterly Almanac. The $16$ items include voting opinions about religion, immigration, crime, education, and other relevant subjects. From a statistical perspective, each item corresponds to a categorical variable with levels \textit{Yes}, \textit{No}, \textit{NA}, the latter in case of a blank ballot.
We apply our method to cluster individuals into groups that are characterized by similarities in the (joint) distribution of the voting opinions. This could help identifying groups of congressmen, possibly belonging to different parties, that are homogeneous in terms of manifest political expressions. In turn, our analysis can establish the existence of factions within a given party, also identifying which political subjects are responsible of such fragmentation.

\begin{figure}[H]
	\centering
	\includegraphics[width=.65\linewidth]{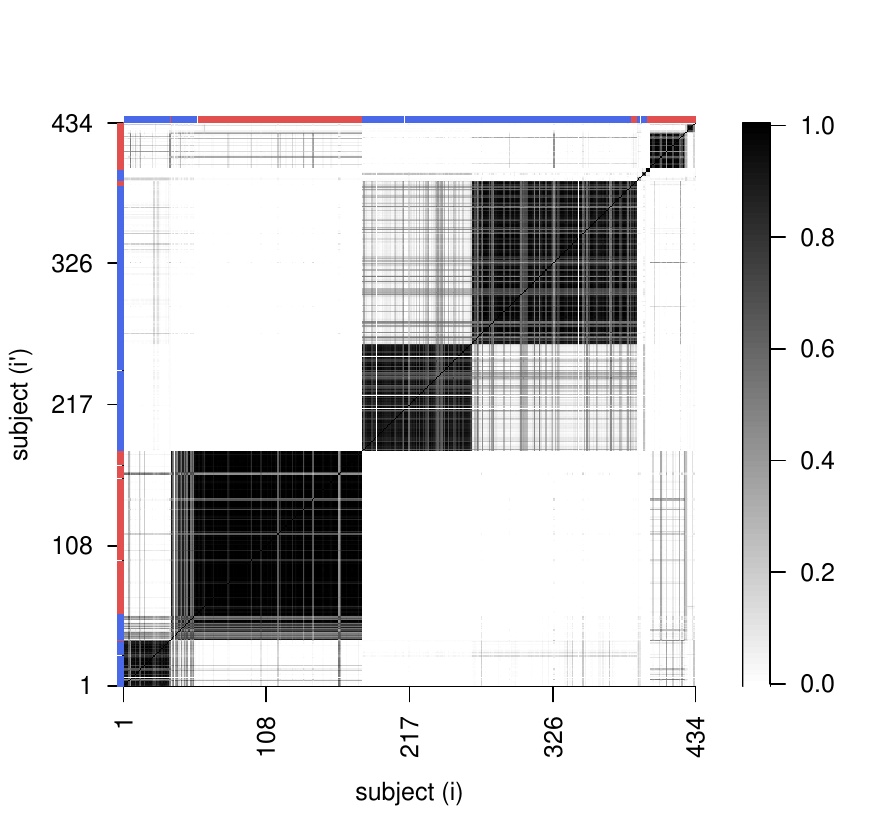}
	\vspace*{-3mm}
	\caption{Voting records. Estimated posterior similarity matrix, with subjects arranged according to the estimated clustering partition, and colors representing their party memberships, blue and red for democrats and republicans respectively.}
	\label{fig:voting:psm}
\end{figure}

Our algorithm is implemented for $S=50000$ iterations after a burn in period of $10000$ draws, 
under the same prior hyperparameter settings of Section \ref{sec:Application:fibrosis}.
Our clustering results are summarized in the heatmap of Figure \ref{fig:voting:psm}, which represents the estimated posterior similarity matrix,
and where subjects have been arranged according to a point estimate of the clustering partition obtained \emph{via} the expected VI criterion. 
%with the $(i,i')$-cell corresponding to the probability in Equation \eqref{eq:post:prob:same:cluster} computed for subjects $i,i'\in\{1,\dots,n\}$.
In addition, colors reported at the borders of the matrix label each congressman according to the party membership, namely democrat (in blue) or republican (in red).

\begin{figure}[H]
    \centering
    \includegraphics[width=0.49\textwidth]{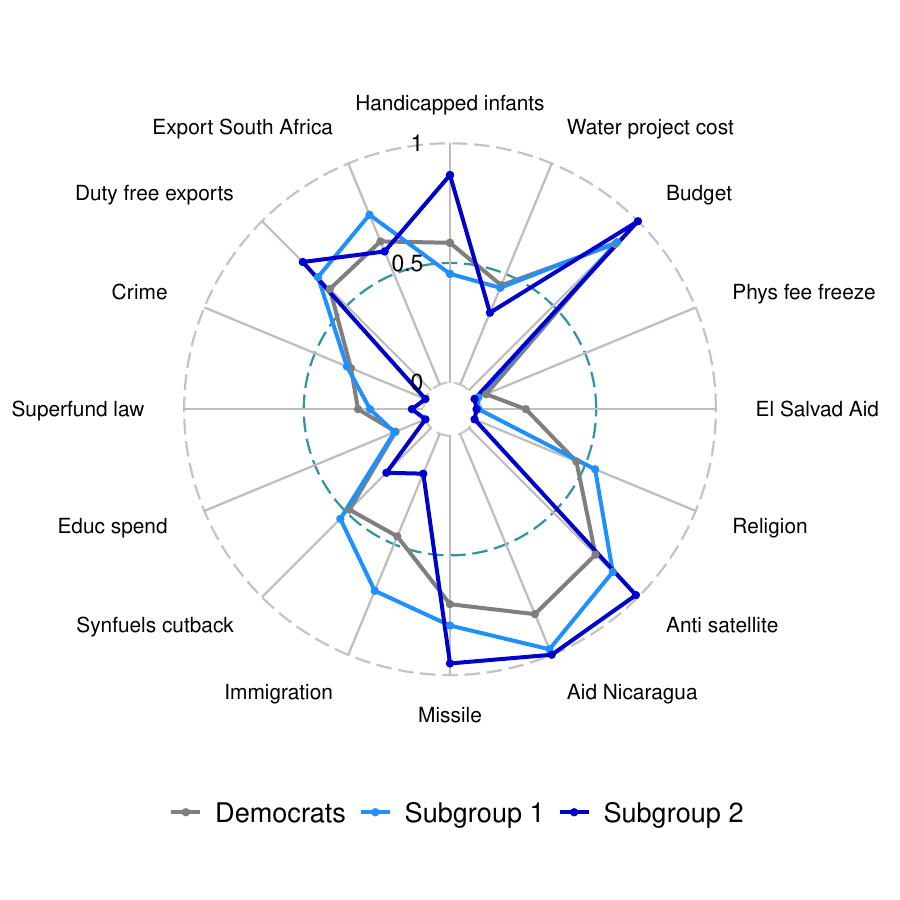}
    \hfill
    \includegraphics[width=0.49\textwidth]{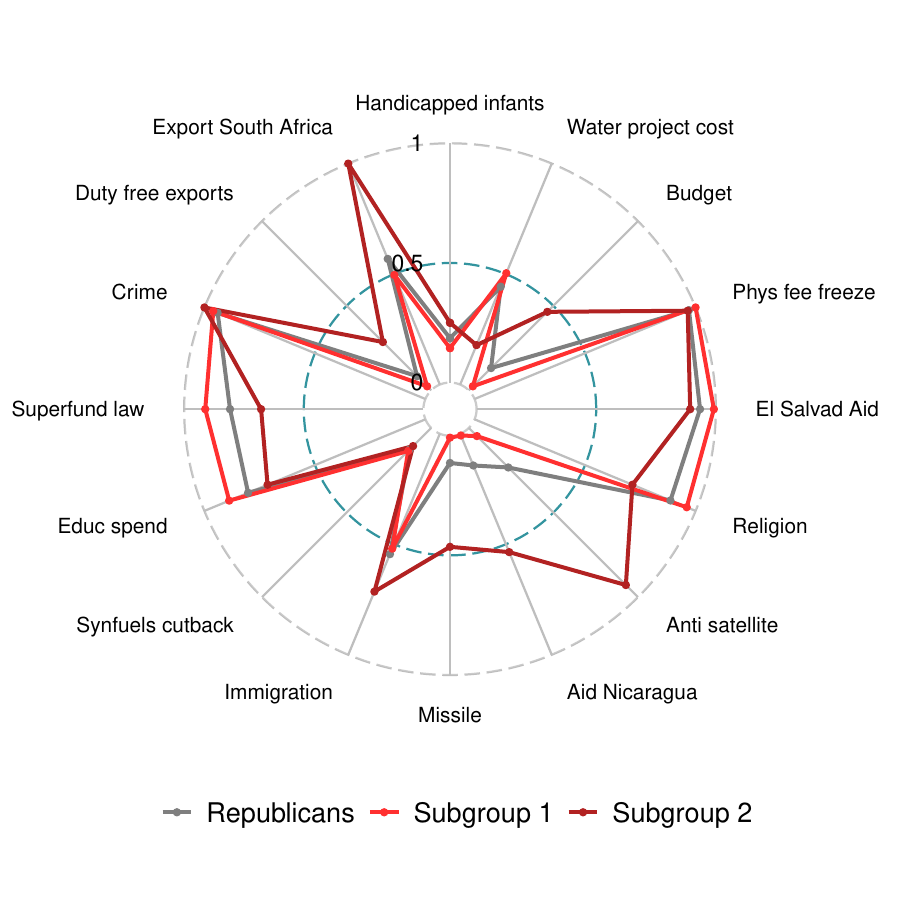}
    \caption{Voting records. Spider plots for the comparison of the party-wide average support proportions (in gray) for each of the $16$ items, with those obtained from the two main party-sub-groups identified by our estimated clustering (colored dots and lines). The two plots refer to democrats (left) and republicans (right).}
    \label{fig:spider:plots}
\end{figure}

The clustering point estimate reveals five main clusters, of which three having a large size (145, 125 and 83 individuals respectively) and thus covering the $81\%$ of the units. Other two smaller clusters are instead moderate-sized, with 36 and 28 observations. Two of the three largest ones collect the majority of the democrats, which are therefore separated into two main sub-groups, while in the third one almost all the included congressmen are republican. We further investigate the role played by each of the $q=16$ items in the identification of the above clustering structure.
To this end, we first group the $n$ individuals according to the two parties.
For each party and for each variable $X_j$, we compute the \emph{party-wide average support}, i.e.~the proportion of \textit{Yes} votes for item $j$ among congressman belonging to the same party.
Then, for both democrats and republicans, we compute the same proportion within each sub-group, identified as the intersection of the party with one of the estimated clusters.
%\gray We then compare these proportions with those obtained by further subdividing Democrats and Republicans according to the point clustering. Differences in issue preferences emerge. In particular, for both parties, the cluster containing the majority of members closely reflects the overall party stance, whereas smaller subgroups display voting patterns that differ from the majority. \black
A comparison of such proportions for the two largest party-sub-groups (labeled as Subgroup 1 and Subgroup 2) is provided in the spider plots of Figure \ref{fig:spider:plots}, corresponding to democrats (left) and republicans (right). The party-wide average support proportions are included as gray dots joined by gray lines. The sub-group proportions are instead reported as colored points and lines.
In both cases, a few differences between sub-groups w.r.t.~the average party proportions emerge. In particular, the two subgroups identified within the democrats are characterized by different opinions regarding immigration policies, with support proportions around $70\%$ and just $20\%$ in Subgroups 1 and 2 respectively.
On the other hand, heterogeneity between the two main republican sub-groups are mostly related to different opinions regarding missile and anti satellite programs.

\section{Conclusion}
\label{sec:Discussion}

We proposed a novel methodology for model-based clustering of categorical data, based on a Dirichlet process mixture of graphical models. A distinctive feature of our model lies in the introduction of a dependence (graphical) structure characterizing the joint distribution of variables within each mixture component.
This provides an effective way to capture heterogeneous dependence patterns between variables, which are common to emerge in several real-world applications. Importantly, these dependencies can vary across latent groups, a feature which enhances the clustering process, particularly w.r.t.~methods developed under the assumption of conditional independence between variables.
Concerns that such an assumption may be unrealistic in applied contexts have been raised in the recent literature, and addressed within the Bayesian framework by \citet{Malsiner:et:al:2024} through a two-layer mixture of finite mixtures latent class model. In contrast to their approach, our methodology explicitly models a dependence structure between variables, which is fully learned from the data under a unified Bayesian framework. Ultimately, learning such dependence structure is of interest in several applications, as for instance the analysis of genomic data based on allelic variants presented in Section \ref{sec:Application}.
We complement our model development with a Markov chain Monte Carlo scheme for posterior inference, and assess its performance in simulation studies. Here, comparisons with alternative methods reveal the benefits of a graphical model-based approach. %Results reveal the benefits of a graphical-model based clustering approach with respect to alternative methods assuming conditional independence between variables.

While offering substantial flexibility, our modeling framework also poses computational challenges that are specifically related to the proposed MCMC scheme. In particular, our algorithm requires an exploration of the graph space, which grows more than exponentially with the number of variables. Future work will therefore focus on the development of scalable strategies for posterior inference, potentially relying on approximate methods. In this vein, a promising extension could be the adoption of variational Bayes approaches, which have demonstrated promising performances for both inference of infinite mixtures as well as of graphical models; see respectively \cite{BealGhahramani2006VB} and \cite{BleiJordan2006VDPM}. Under this framework, the posterior distribution would be approximated by the member of a tractable family which is closest to the true posterior in terms of Kullback-Leibler divergence.

In both applications of Section \ref{sec:Application}, subjects are already structured into groups, such as case and control patients or democrat and republican congressmen, although this information is not accounted for by our model.
In a more flexible setting, one could introduce this hierarchical structure, particularly to allow sharing of information across groups, and at the same time allowing for common graphical structures among individuals belonging to different groups.
From a modeling perspective, this setting could be employed by relying on hierarchical mixtures, such as based, in a Bayesian nonparametric framework, on the Hierarchical Dirichlet Process \citep{Teh:et:al:2006}. 
%This provides an effective way to construct such models by enabling groups to share mixture components. This is achieved by assigning each group a random probability measure drawn from a Dirichlet Process, whose base measure is in turn random and distributed accordingly to another Dirichlet Process.

\section*{Supplementary Materials}

Supplementary Material is organized into two sections. Section 1 provides details regarding the computation of the marginal likelihood of a decomposable graph. Section 2 instead provides technical results regarding our MCMC scheme, including the proofs of Propositions 4.1 and 4.2, and the construction of the proposal distribution over the space of decomposable undirected graphs.
R code implementing our methodology is publicly available at \url{https://anonymous.4open.science/status/GraphicalModel-BasedClustering-37F6}. The same repository contains the analyzed datasets.

\section*{Acknowledgments}

We thank Lucia Paci (UCSC) for helpful discussion during the empirical validations of our methodology.
Work carried out within MUR-PRIN grant 2022 SMNNKY-CUP J53D23003870008, funded by the European Union Next Generation EU. The views and opinions expressed are only those of the Authors and do not necessarily reflect those of the European Union or the European Commission. Neither the European Union nor the European Commission can be held responsible for them. Partial support from UCSC (D1 research grant) is also acknowledged by F.C.

\bibliographystyle{biometrika}
\bibliography{biblio_2}

\begin{thebibliography}{30}
\expandafter\ifx\csname natexlab\endcsname\relax\def\natexlab#1{#1}\fi
\expandafter\ifx\csname url\endcsname\relax
  \def\url#1{\texttt{#1}}\fi
\expandafter\ifx\csname urlprefix\endcsname\relax\def\urlprefix{URL }\fi
\providecommand{\eprint}[2][]{\url{#2}}

\bibitem[{Antoniak(1974)}]{Antoniak:1974}
\textsc{Antoniak, C.~E.} (1974).
\newblock Mixtures of {D}irichlet processes with applications to {B}ayesian
  nonparametric problems.
\newblock \textit{Annals of Statistics} 2 1152 -- 1174.

\bibitem[{Argiento \& De~Iorio(2022)}]{Argiento:De:Iorio:2022}
\textsc{Argiento, R.} \& \textsc{De~Iorio, M.} (2022).
\newblock Is infinity that far? {A} {B}ayesian nonparametric perspective of
  finite mixture models.
\newblock \textit{The Annals of Statistics} 50 2641 -- 2663.

\bibitem[{Argiento et~al.(2025)Argiento, Filippi-Mazzola \&
  Paci}]{argientopaci}
\textsc{Argiento, R.}, \textsc{Filippi-Mazzola, E.} \& \textsc{Paci, L.}
  (2025).
\newblock Model-based clustering of categorical data based on the {H}amming
  distance.
\newblock \textit{Journal of the American Statistical Association} 120 1178 --
  1188.

\bibitem[{Beal \& Ghahramani(2006)}]{BealGhahramani2006VB}
\textsc{Beal, M.~J.} \& \textsc{Ghahramani, Z.} (2006).
\newblock Variational {B}ayesian learning of directed graphical models with
  hidden variables.
\newblock \textit{Bayesian Analysis} 1 793 -- 831.

\bibitem[{Blei \& Jordan(2006)}]{BleiJordan2006VDPM}
\textsc{Blei, D.~M.} \& \textsc{Jordan, M.~I.} (2006).
\newblock Variational inference for {D}irichlet process mixtures.
\newblock \textit{Bayesian Analysis} 1 121 -- 143.

\bibitem[{Castelletti \& Peluso(2021)}]{Castelletti:Peluso:2021:CSDA}
\textsc{Castelletti, F.} \& \textsc{Peluso, S.} (2021).
\newblock Equivalence class selection of categorical graphical models.
\newblock \textit{Computational Statistics \& Data Analysis} 164 107304.

\bibitem[{Castelo \& Perlman(2004)}]{Castelo:Perlman:2004}
\textsc{Castelo, R.} \& \textsc{Perlman, M.~D.} (2004).
\newblock Learning essential graph {M}arkov models from data.
\newblock In \textit{Advances in {B}ayesian networks}, vol. 146 of
  \textit{Studies in Fuzziness and Soft Computing}. Springer, Berlin, 255 --
  269.

\bibitem[{Dawid \& Lauritzen(1993)}]{Dawid:Laur:1993}
\textsc{Dawid, A.~P.} \& \textsc{Lauritzen, S.~L.} (1993).
\newblock Hyper {M}arkov laws in the statistical analysis of decomposable
  graphical models.
\newblock \textit{The Annals of Statistics} 21 1272 -- 1317.

\bibitem[{Escobar \& West(1994)}]{escobar}
\textsc{Escobar, M.} \& \textsc{West, M.} (1994).
\newblock Bayesian density estimation and inference using mixtures.
\newblock \textit{Journal of the American Statistical Association} 90 577 --
  588.

\bibitem[{Ferguson(1973)}]{Ferguson:1973}
\textsc{Ferguson, T.~S.} (1973).
\newblock A {B}ayesian analysis of some nonparametric problems.
\newblock \textit{The Annals of Statistics} 1 209 -- 230.

\bibitem[{Geiger \& Heckerman(1997)}]{Geiger:Heckerman:1997}
\textsc{Geiger, D.} \& \textsc{Heckerman, D.} (1997).
\newblock A characterization of the {D}irichlet distribution through global and
  local parameter independence.
\newblock \textit{Annals of Statistics} 25 1344 -- 1369.

\bibitem[{Heckerman et~al.(1995)Heckerman, Geiger \&
  Chickering}]{Heckerman:et:al:1995}
\textsc{Heckerman, D.}, \textsc{Geiger, D.} \& \textsc{Chickering, D.~M.}
  (1995).
\newblock Learning {B}ayesian networks: {T}he combination of knowledge and
  statistical data.
\newblock \textit{Machine Learning} 20 197 -- 243.

\bibitem[{Huang(1998)}]{kmodes}
\textsc{Huang, Z.} (1998).
\newblock Extensions to the k-{M}eans algorithm for clustering large data sets
  with categorical values.
\newblock \textit{Data Mining and Knowledge Discovery} 2 283 -- 304.

\bibitem[{Kalisch \& Bühlmann(2007)}]{Kalisch:Buhlmann:2007}
\textsc{Kalisch, M.} \& \textsc{Bühlmann, P.} (2007).
\newblock Estimating high-dimensional directed acyclic graphs with the
  {PC}-{A}lgorithm.
\newblock \textit{Journal of Machine Learning Research} 8 613 -- 636.

\bibitem[{Kerem et~al.(1989)Kerem, Rommens, Buchanan, Markiewicz, Cox,
  Chakravarti, Buchwald \& Tsui}]{Kerem:et:al:89}
\textsc{Kerem, B.}, \textsc{Rommens, J.}, \textsc{Buchanan, J.},
  \textsc{Markiewicz, D.}, \textsc{Cox, T.}, \textsc{Chakravarti, A.},
  \textsc{Buchwald, M.} \& \textsc{Tsui, L.} (1989).
\newblock Identification of the cystic fibrosis gene: Genetic analysis.
\newblock \textit{Science} 245 1073 -- 1080.

\bibitem[{Lauritzen(1996)}]{Laur:1996}
\textsc{Lauritzen, S.~L.} (1996).
\newblock \textit{Graphical Models}.
\newblock Oxford University Press.

\bibitem[{Linzer \& Lewis(2011)}]{linzer}
\textsc{Linzer, D.~A.} \& \textsc{Lewis, J.~B.} (2011).
\newblock {poLCA: An R package for polytomous variable latent class analysis}.
\newblock \textit{Journal of Statistical Software} 42 1 -- 29.

\bibitem[{Liu et~al.(2001)Liu, Sabatti, Teng, Keats \&
  Risch}]{Sabatti:et:al:2001}
\textsc{Liu, J.~S.}, \textsc{Sabatti, C.}, \textsc{Teng, J.}, \textsc{Keats,
  B.~J.} \& \textsc{Risch, N.} (2001).
\newblock Bayesian analysis of haplotypes for linkage disequilibrium mapping.
\newblock \textit{Genome research} 11 1716 -- 1724.

\bibitem[{MacQueen(1967)}]{MacQueen:1967}
\textsc{MacQueen, J.} (1967).
\newblock Some methods for classification and analysis of multivariate
  observations.
\newblock In \textit{Volume 1: Statistics}, Proceedings of the Fifth Berkeley
  Symposium on Mathematical Statistics and Probability. University of
  California Press, 281 -- 297.

\bibitem[{Madigan \& York(1995)}]{Madigan:York:1995}
\textsc{Madigan, D.} \& \textsc{York, J.} (1995).
\newblock Bayesian graphical models for discrete data.
\newblock \textit{International Statistical Review} 63 215 -- 232.

\bibitem[{Malsiner-Walli et~al.(2025)Malsiner-Walli, Grün \&
  Frühwirth-Schnatter}]{Malsiner:et:al:2024}
\textsc{Malsiner-Walli, G.}, \textsc{Grün, B.} \&
  \textsc{Frühwirth-Schnatter, S.} (2025).
\newblock Without pain -- {C}lustering categorical data using a {B}ayesian
  mixture of finite mixtures of latent class analysis models.
\newblock \textit{Advances in Data Analysis and Classification} .

\bibitem[{Neal(2000)}]{Neal:2000}
\textsc{Neal, R.~M.} (2000).
\newblock Markov chain sampling methods for {D}irichlet process mixture models.
\newblock \textit{Journal of Computational and Graphical Statistics} 9 249 --
  265.

\bibitem[{Rodr{\'i}guez et~al.(2011)Rodr{\'i}guez, Lenkoski \&
  Dobra}]{Rodriguez:et:al:2009}
\textsc{Rodr{\'i}guez, A.}, \textsc{Lenkoski, A.} \& \textsc{Dobra, A.} (2011).
\newblock Sparse covariance estimation in heterogeneous samples.
\newblock \textit{Electronic Journal of Statistics} 5 981 -- 1014.

\bibitem[{Russell \& Norvig(2016)}]{Russell:2016}
\textsc{Russell, S.~J.} \& \textsc{Norvig, P.} (2016).
\newblock \textit{Artificial Intelligence: a Modern Approach}.
\newblock Pearson.

\bibitem[{Schaid et~al.(2018)Schaid, Chen \& Larson}]{Schaid:et:al:2018}
\textsc{Schaid, D.}, \textsc{Chen, W.} \& \textsc{Larson, N.} (2018).
\newblock From genome-wide associations to candidate causal variants by
  statistical fine-mapping.
\newblock \textit{Nature Reviews Genetics} 19 491 -- 504.

\bibitem[{Scutari \& Denis(2021)}]{Scutari:Denis:R:2021}
\textsc{Scutari, M.} \& \textsc{Denis, J.-B.} (2021).
\newblock \textit{Bayesian Networks: {W}ith Examples in {R} (2nd ed.)}.
\newblock Chapman and Hall/CRC.

\bibitem[{Spirtes et~al.(2000)Spirtes, Glymour \&
  Scheines}]{Spir:Glym:Sche:2000}
\textsc{Spirtes, P.}, \textsc{Glymour, C.} \& \textsc{Scheines, R.} (2000).
\newblock \textit{Causation, Prediction and Search (2nd edition).}
\newblock Cambridge, MA: The MIT Press.

\bibitem[{Teh et~al.(2006)Teh, Jordan, Beal \& Blei}]{Teh:et:al:2006}
\textsc{Teh, Y.~W.}, \textsc{Jordan, M.~I.}, \textsc{Beal, M.~J.} \&
  \textsc{Blei, D.~M.} (2006).
\newblock Hierarchical {D}irichlet processes.
\newblock \textit{Journal of the American Statistical Association} 101
  1566--1581.

\bibitem[{Wade \& Ghahramani(2018)}]{wade:2018}
\textsc{Wade, S.} \& \textsc{Ghahramani, Z.} (2018).
\newblock Bayesian cluster analysis: {P}oint estimation and credible balls
  (with discussion).
\newblock \textit{Bayesian Analysis} 13 559 -- 626.

\bibitem[{White \& Murphy(2014)}]{White:Murphy:BayesLCA}
\textsc{White, A.} \& \textsc{Murphy, T.~B.} (2014).
\newblock {BayesLCA: An R package for Bayesian latent class analysis}.
\newblock \textit{Journal of Statistical Software} 61 1 -- 28.

\end{thebibliography}


\begin{thebibliography}{0}
\expandafter\ifx\csname natexlab\endcsname\relax\def\natexlab#1{#1}\fi
\expandafter\ifx\csname url\endcsname\relax
  \def\url#1{\texttt{#1}}\fi
\expandafter\ifx\csname urlprefix\endcsname\relax\def\urlprefix{URL }\fi
\providecommand{\eprint}[2][]{\url{#2}}

\end{thebibliography}

\end{document}

\title{Supplementary Material to \\ Graphical model-based clustering of categorical data}
%\author[1]{Laura Ferrini \thanks{l.ferrini2@campus.unimib.it}}
%\author[2]{Federico Castelletti \thanks{federico.castelletti@unicatt.it}}
%\affil[1,2]{Department of Statistical Sciences, Universit\`{a} Cattolica del Sacro Cuore, Milan}
%\affil[1]{Department of Economics, Management and Statistics, Universit\`{a} degli Studi di Milano-Bicocca, Milan}

\author{}
\date{}

\maketitle

%\maketitle
%\noindent

This Supplementary Material is organized into two sections. Section 1 provides details relative to the computation of the marginal likelihood of a categorical decomposable graphical model. Section 2 provides technical results regarding our MCMC scheme, specifically the closed-form expressions of the prior and posterior predictive distributions involved in the full conditional of cluster indicators, and details regarding the proposal distribution implemented in the Metropolis Hastings update of a decomposable Undirected Graph (UG).
%\section{Introduction}
%\label{sec:Introduction}

\vspace{0.4cm}

\section{Marginal likelihood of decomposable UG}

Let $\bX$ be the $(n, q)$ data matrix collecting $n$ $q$-variate observations $\bx^{(i)}=(x_1^{(i)},\dots,x_q^{(i)})^\top$, $i=1,\dots,n$, from $X_1,\dots,X_q$ categorical variables.
Under a decomposable UG $\G$ having sets of cliques and separators $\mathcal{C}$ and $\mathcal{S}$ respectively, the likelihood function factorizes as
\be
p(\bX\g\btheta,\G)=
\frac
{\prod_{C \in \mathcal{C}}p\left(\bX_{C}\g\btheta_C\right)}
{\prod_{S \in \mathcal{S}}p\left(\bX_{S}\g\btheta_S\right)},
\ee
where $\bX_S$ is the sub-matrix of $\bX$ with columns indexed by $S \subseteq \{1,\dots,q\}$.
Under a Hyper-Dirichlet law, the prior on $\btheta$ also admits the factorization
\be
p(\btheta\g\G)=
\frac
{\prod_{C \in \mathcal{C}}p\left(\btheta_C\right)}
{\prod_{S \in \mathcal{S}}p\left(\btheta_S\right)}.
\ee
%where $\mathcal{C}$ and $\mathcal{S}$ are respectively the set of cliques and separators of $\G$.
As a consequence, the marginal data distribution $m(\bX\g\G)=\int p(\bX\g\btheta,\G)p(\btheta\g\G)\, d\btheta$ can be written as
\ben
\label{eq:marginal:likelihood:factorization}
m(\bX\g\G)=
\frac
{\prod_{C \in \mathcal{C}}m\left(\bX_{C}\right)}
{\prod_{S \in \mathcal{S}}m\left(\bX_{S}\right)}.
\een
Furthermore, under the assumptions
\be
p(\bX_S\g\btheta_S) = \prod_{x_S\in\X_S} \left\{\theta^S_{x_S}\right\}^{n_{x_S}^S}
\quad\quad
p(\btheta_S) =
\frac{\Gamma\big(\sum_{x_c\in\X_S} a^S_{x_S} \big)}
{\prod_{x_S\in\X_S}\Gamma\big(a^S_{x_S}\big)}
\prod_{x_S\in\X_S}\left\{\theta^S_{x_S}\right\}^{a^S_{x_S}-1},
\ee
respectively a categorical distribution for $\bX_S$ and a Dirichlet prior on $\btheta_S$ (see also Section 2 of the main text), we obtain, for generic $S \subseteq \{1,\dots,q\}$,
\be
m(\bX_S) = \frac{\Gamma\left(\sum_{x_S \in \X_S}a_{x_S}^S\right)}
{\Gamma\left(\sum_{x_S \in \X_S}\left(a_{x_S}^S+n_{x_S}^S\right)\right)}
\prod_{x_S \in \X_S}\frac{\Gamma\left(a_{x_S}^S+n_{x_S}^S\right)}
{\Gamma\left(a_{x_S}^S\right)}.
\ee
%where in particular $\X_S$ is the set of levels of variables in $S$.
In addition, under the default choice $a_{x_S}^S=\frac{a}{|\X_S|}$ the previous expression can be simplified as
\ben
\label{eq:marginal:S}
m(\bX_S) = \frac{\Gamma\left(a\right)}
{\Gamma\left(a+n\right)}
\prod_{x_S \in \X_S}\frac{\Gamma\left(a_{x_S}^S+n_{x_S}^S\right)}
{\Gamma\left(a_{x_S}^S\right)}.
\een

\vspace{0.4cm}

\section{MCMC scheme}

Our MCMC scheme consists of a Gibbs sampling scheme targeting the marginal posterior of cluster indicators, graphs and concentration parameter
$
p\left(\{\xi_i\}_{i = 1}^n, \{\mathcal{G}_k\}_{k =1}^K, \alpha, K \g \bX \right)
$,
where $K$ is the number of non-empty clusters, i.e.~unique values among the $\xi_i$'s.
The algorithm is based on the following iterative steps:
\begin{enumerate}	
	\item Sample the cluster indicators $\{\xi\}_{i = 1}^n$ from their full conditional $p(\{\xi_i\}_{i = 1}^n \g \textnormal{rest})$;
	\item Sample the graphs $\{\mathcal{G}_k\}_{k=1}^K$ from their full conditional $p(\{\mathcal{G}_k\}_{k =1}^K \g \textnormal{rest})$;
	\item Sample the DP concentration parameter $\alpha$ from its full conditional $p(\alpha\g \textnormal{rest})$.
\end{enumerate}
In the following subsections we provide some technical details regarding the first two steps. 

\subsection{Full conditional of cluster indicators}

The full conditional of cluster indicator $\xi_i$ is given by:
\ben
\label{eq:full:cond:cluster}
	\begin{aligned}
		p\big(\xi_i = k \g \{\bx^{(l)} : l \neq i, \xi_l = k\}, \G_k\big) \propto
		\begin{cases}
			\,n^{-i}_k \, p\big(\bx^{(i)} \g  \{\bx^{(l)} : l \neq i, \xi_l = k\}, \G_k\big) & k = 1,\dots,K \\
			\,\alpha \, p\big(\bx^{(i)} \g \G_{k}\big) & k = K + 1,
		\end{cases}
	\end{aligned}
\een
where $n_k^{-i}$ is the number of observations assigned to cluster $k$, excluding $\bx^{(i)}$; see also Section 4 of our main paper.

\vspace{0.5cm}

The first case of Equation \eqref{eq:full:cond:cluster} considers the case of a non-empty cluster $(k=1,\dots,K)$. In this case, the full conditional probability of $\xi_i=k$ is proportional to the posterior predictive probability of $\bx^{(i)}$ conditionally on all other observations currently assigned to cluster $k$. 
The second case of Equation \eqref{eq:full:cond:cluster} instead refers to a new empty cluster, say $K+1$, with no observations that are currently assigned to it. In such a case, the full conditional of $\xi_i$ is proportional to the prior predictive distribution evaluated at $\bx^{(i)}$.
Both quantities admit closed-form expressions that are summarized in Propositions 4.1 and 4.2 of the main text and for which we provide below detailed proofs.

\subsection*{Proof of Proposition 4.1: \\Posterior predictive distribution - non-empty cluster}
The posterior predictive distribution of $\bx^{(i)}$ for a non-empty cluster $k$ can be equivalently written as the ratio of marginal likelihoods
\be
p\big(\bx^{(i)} \g  \bX^{(k)-i}, \G_k\big)
=
\frac{m(\bX^{(k)+i}\gbig\G_k)}{m(\bX^{(k)-i}_{\white.\black}\gbig\G_k)},
\ee
where $\bX^{(k)+i}=\{\bx^{(l)}: \xi_l = k\}\cup \bx^{(i)}$ and $\bX^{(k)-i}=\{\bx^{(l)}: l \neq i, \xi_l = k\}$. Furthermore, each marginal likelihood factorizes according to $\G_k$ as in Equation \eqref{eq:marginal:likelihood:factorization}.
Therefore, we can write
\ben
\label{eq:predictive:decomposable}
p\big(\bx^{(i)} \g  \bX^{(k)-i}, \G_k\big)
\,=\,
%\frac{m\big(\bX^{(k)}\gbig\G_k\big)}{m\big(\bX^{(k)-i}_{\white.\black}\gbig\G_k\big)} \,=\,
\frac{\prod_{C \in \mathcal{C}_k}
m\left(\bX_C^{(k)+i}\right)\big/m\left(\bX_C^{(k)-i}\right)}
{\prod_{S \in \mathcal{S}_k}
m\left(\bX_S^{(k)+i}\right)\big/m\left(\bX_S^{(k)-i}\right)},
\een
where $\mathcal{C}_k$ and $\mathcal{S}_k$ are respectively the set of cliques and separators of $\G_k$.
Now consider the generic term $m\big(\bX_S^{(k)+i}\big)\big/m\big(\bX_S^{(k)-i}\big)$ where to ease the notation we omit index $k$ from $\bX^{(k)}$ since we reason conditionally on a fixed $k$. Because of \eqref{eq:marginal:S} we can write
\ben
\label{eq:marginal:ratio:clusters}
\frac{m\big(\bX_S^{+i}\big)}
{m\big(\bX_S^{-i}\big)}
=
\frac{\Gamma\left(a+n^{-i}\right)}
{\Gamma\big(a+n^{+i}\big)}
\prod_{x_S \in \X_S}\frac{\Gamma\left(a_{x_S}^S+n_{x_S}^{S}(+i)\right)}
{\Gamma\left(a_{x_S}^S+n_{x_S}^{S}(-i)\right)},
\een
where $n_{x_S}^{S}(+i)$ and $n_{x_S}^{S}(-i)$ are the counts associated with configuration $x_S$ as obtained from $\bX^{+i}_S$ and $\bX_S^{-i}$ respectively, $n^{+i}$ and $n^{-i}$ the sample sizes of the two data matrices, which do not depend on $S$.
Furthermore, notice that $\bX^{+i}_S$ and $\bX_S^{-i}$ only differ by one observations, $\bx^{(i)}$, for which we have $\bx_S^{(i)}$ as the observed level of variables in $S$; let for simplicity $\bx^{(i)}=\tilde x$.
Accordingly, all the $x_S$-terms in \eqref{eq:marginal:ratio:clusters} will cancel out except for those corresponding to $x_S = \widetilde x_S$, so that we can simplify \eqref{eq:marginal:ratio:clusters} as
\be
\label{eq:marginal:ratio:clisters}
\frac{m\big(\bX_S^{+i}\big)}
{m\big(\bX_S^{-i}\big)}
=
\frac{\Gamma\left(a+n^{-i}\right)}
{\Gamma\big(a+n^{+i}\big)}\cdot
\frac
{\Gamma\left(a_{\widetilde x_S}^S+n_{\widetilde x_S}^{S}(+i)\right)}
{\Gamma\left(a_{\widetilde x_S}^S+n_{\widetilde x_S}^{S}(-i)\right)}.
\ee
We now distinguish two cases:
\begin{itemize}
	\item[$(a)$] $\xi_i = k$, i.e.~subject $i$ belongs to cluster $k$;
	\item[$(b)$] $\xi_i \ne k$, i.e.~subject $i$ does not belong to cluster $k$.
\end{itemize}
In case $(a)$ we have $n^{-i} = n-1$, $n^{+i} = n$ and $n_{\widetilde x_S}^{S}(-i)=n_{\widetilde x_S}^{S}-1$. Therefore, we can write
\be
\frac{m\big(\bX_S^{+i}\big)}
{m\big(\bX_S^{-i}\big)}
=
\frac
{a_{\widetilde x_S}^S+n_{\widetilde x_S}^{S}-1}
{a+n-1^{\white '}},
\ee
using the fact that $\Gamma(c)=(c-1)\Gamma(c-1)$, for any constant $c>0$.
In case $(b)$ instead,
$n^{-i} = n$, $n^{+i} = n+1$ and $n_{\widetilde x_S}^{S}(+i)=n_{\widetilde x_S}^{S}+1$. Accordingly, we get
\be
\frac{m\big(\bX_S^{+i}\big)}
{m\big(\bX_S^{-i}\big)}
=
\frac
{a_{\widetilde x_S}^S+n_{\widetilde x_S}^{S}}
{a+n^{\white '}}.
\ee
By combining together the two cases, we can write
\be
\frac{m\big(\bX_S^{+i}\big)}
{m\big(\bX_S^{-i}\big)}
=
\frac
{a_{\widetilde x_S}^S+n_{\widetilde x_S}^{S}-\mathbbm{1}\left\{\xi_i=k\right\}}
{a+n-\mathbbm{1}\left\{\xi_i=k\right\}^{\white '}},
\ee
which provides a general formula for the generic $S$- and $C$-terms in Equation
\eqref{eq:predictive:decomposable}.
The latter can thus be written as
\ben
\label{eq:predictive:non:empty}
p\big(\bx^{(i)} \g \bX^{(k)-i}, \G_k\big)
\,=\,
\big\{
a+n_k-\mathbbm{1}\left\{\xi_i=k\right\}
\big\}^{|\mathcal{S}_k|-|\mathcal{C}_k|}\cdot
\frac{\prod_{C \in \mathcal{C}_k} \Big(a_{\bx^{(i)}_C}^C+\prescript{}{k}n_{\bx^{(i)}_C}^{C}-\mathbbm{1}\left\{\xi_i=k\right\}\Big)}
{\prod_{S \in \mathcal{S}_k} \Big(a_{\bx^{(i)}_S}^S+\prescript{}{k}n_{\bx^{(i)}_S}^{S}-\mathbbm{1}\left\{\xi_i=k\right\}\Big)},
\een
now restoring the dependence of the counts (and hyperparameters) from $k$ and $i$.

\subsection*{Proof of Proposition 4.2: \\ Prior predictive distribution - empty cluster}
The prior predictive distribution of $\bx^{(i)}$ factorizes according to the decomposable UG $\G_k$ as
\ben
\label{eq:marginal:decomposable:new:cluster}
p\big(\bx^{(i)} \g \G_k\big)=
\frac
{\prod_{C \in \mathcal{C}_k}m\left(\bx_{C}^{(i)}\right)}
{\prod_{S \in \mathcal{S}_k}m\left(\bx_{S}^{(i)}\right)};
\een
see also Equation \eqref{eq:marginal:likelihood:factorization}.
Each generic term $m\big(\bx_S^{(i)}\big)$, which can be written as in \eqref{eq:marginal:S}, is now based on a sample of size one, namely $\bx^{(i)}$.
Accordingly, all counts $n_{x_S}^S$ in \eqref{eq:marginal:S} will be zero except for the one corresponding to $x_S=\bx_S^{(i)}$ which will be instead equal to one. Therefore, it is easy to show, again using properties of Gamma functions, that 
$
m\big(\bx_{S}^{(i)}\big) = a_{\bx^{(i)}_S}^S/a.
$
Finally, Equation \eqref{eq:marginal:decomposable:new:cluster} can be written as
\ben
\label{eq:marginal:empty}
p\big(\bx^{(i)} \g \G_k\big)
\,=\,
\big\{a\big\}^{|\mathcal{S}_k|-|\mathcal{C}_k|}
\cdot
\frac
{\prod_{C \in \mathcal{C}_k}a_{\bx^{(i)}_C}^C}
{\prod_{S \in \mathcal{S}_k}a_{\bx^{(i)}_S}^S}.
\een

\subsection{Full conditional of graph $\mathcal{G}_k$}
The full conditional distribution of the decomposable UG $\G_k$ is not available in closed form. Accordingly, the graph update is performed through a Metropolis Hastings (MH) step applied to each cluster $k$ which at the current MCMC iteration is not empty. To simplify the notation, we omit the subscript $k$ from $\mathcal{G}_k$ and $\bX^{(k)}$, so that the dependence from cluster $k$ is tacitly assumed. An MH update requires a proposal distribution $q(\tilde{\mathcal{G}} \g \mathcal{G})$ determining the transition from the current graph $\mathcal{G}$ to a new graph $\tilde{\mathcal{G}}$.
To build such proposal, we consider two types of moves that can be applied to graph $\G$: edge \emph{deletion} and edge \emph{insertion}. These moves can be applied to distinct pairs of nodes $u\neq v$, with $u,v\in\{1,\dots,q\}$ provided that the resulting graph (i.e.~the graph obtained by applying the move to $\G$) is a decomposable graph. For a given $\G$, we thus obtain a collection of \emph{valid} operations corresponding to a set of decomposable graphs that we denote by $\mathcal{O}_{\mathcal{G}}$. The proposal distribution $q(\tilde{\mathcal{G}} \g \mathcal{G})$ then assigns uniform probabilities to the elements in $\mathcal{O}_{\mathcal{G}}$; consequently, the proposal probability of $\tilde\G \in \mathcal{O}_{\mathcal{G}}$ is given by
\[
q(\tilde{\mathcal{G}} \g \mathcal{G}) = \frac{1}{\lvert \mathcal{O}_{\mathcal{G}} \rvert}.
\]
Finally, in an MH scheme, graph $\tilde{\mathcal{G}}$ is accepted with probability
\[
\alpha_{\mathcal{G}} =
\min\left\{
1,\,
\frac{m(\boldsymbol{X} \g \tilde{\mathcal{G}})}{m(\boldsymbol{X} \g \mathcal{G})} \cdot
\frac{p(\tilde{\mathcal{G}})}{p(\mathcal{G})} \cdot
\frac{q(\mathcal{G} \g \tilde{\mathcal{G}})}{q(\tilde{\mathcal{G}} \g \mathcal{G})}
\right\},
\]
where $p(\G)$ is the prior on graph $\G$ (Section 3.2 of our main text), while the closed-form expression of the marginal likelihood $m(\boldsymbol{X} \g \mathcal{G})$ is given in Equations \eqref{eq:marginal:likelihood:factorization} and \eqref{eq:marginal:S}.

%\bibliographystyle{biometrika}
%\bibliography{biblio_2}